\documentclass[twocolumn,aps,pre]{revtex4-1}

\usepackage{amsmath,amssymb}
\usepackage{graphicx}
\usepackage{placeins}
\usepackage{rotating}
\usepackage{xspace}
\usepackage{color}

\newcommand{\Gasper}{Ga\v{s}per Tka\v{c}ik}

\newcommand{\nn}{\nonumber}
\newcommand{\myvec}[1]{\mathbf{#1}}
\newcommand{\la}{\langle}
\newcommand{\ra}{\rangle}
\makeatletter
\newcommand{\vast}{\bBigg@{3}}
\newcommand{\Vast}{\bBigg@{4}}
\newcommand{\giant}{\bBigg@{5}}
\newcommand{\Giant}{\bBigg@{6}}
\makeatother
\newcommand{\ls}{\delta} 
\newcommand{\hoprate}{h} 
\newcommand{\dg}{\delta g}
\newcommand{\mean}{\bar g} 
\newcommand{\cc}{C} 
\newcommand{\DoverLS}{\frac{D}{\ls^2}}
\newcommand{\rmax}{r_{\rm max}} 
\newcommand{\Nmax}{N_{\rm max}}
\newcommand{\cmax}{c_{\rm max}}
\newcommand{\fromNNof}[1]{\in N(#1)}
\newcommand{\notfromNNof}[1]{\notin N(#1)}

\newcommand{\Noise}{\mathcal{N}}
\newcommand{\NNoise}{\mathcal{N}}
\newcommand{\GNoise}{\mathcal{G}}
\newcommand{\DNoise}{\mathcal{D}}

\newcommand{\Drosophila}{\textit{Drosophila}\xspace}

\begin{document}

\title{Optimizing information flow in small genetic networks.\\ IV. Spatial coupling}
\author{Thomas R. Sokolowski}
\email{thomas.sokolowski@ist.ac.at}
\author{\Gasper}
\email{gasper.tkacik@ist.ac.at}
\affiliation{Institute of Science and Technology Austria, Am Campus 1, A-3400 Klosterneuburg, Austria}

\begin{abstract}
    We typically think of cells as responding to external signals independently by regulating their gene expression levels, yet they often locally exchange information and coordinate. 
    Can such spatial coupling be of benefit for conveying signals subject to gene regulatory noise? 
    Here we extend our information-theoretic framework for gene regulation to spatially extended systems. 
    As an example, we consider a lattice of nuclei responding to a concentration field of a transcriptional regulator (the ``input'') by expressing a single diffusible target gene. 
    When input concentrations are low, diffusive coupling markedly improves information transmission; 
    optimal gene activation functions also systematically change. 
    A qualitatively new regulatory strategy emerges where individual cells respond to the input in a nearly step-like fashion that is subsequently averaged out by strong diffusion. 
    While motivated by early patterning events in the \Drosophila embryo, our framework is generically applicable to spatially coupled stochastic gene expression models.
\end{abstract}

\maketitle

\section{Introduction}
\label{secIntro}
Signals encoded by spatial or temporal concentration profiles of certain proteins play a crucial role in communicating information within and between cells.
Such signals are established and read out by a vast variety of gene regulatory networks.
In this process, however, information flow is limited by the randomness associated with gene regulatory interactions---essentially 
chemical reactions between different species of signaling molecules and DNA---taking place at very low copy numbers \cite{vanKampen}. 
While we increasingly understand distinct strategies of noise control in biological systems \cite{Tsimring2014,Lo2015,Zhang2012,Eldar2004,Eldar2002},
it remains largely unclear how nature orchestrates these strategies to maximize information flow.

A recent proposal takes the idea of ``information flow'' in biological networks seriously, as formalized by Shannon's information theory, 
and hypothesizes that at least some of the structure of genetic regulatory networks could be derived mathematically, 
by maximizing information transmission that such networks can sustain subject to biophysically realistic constraints \cite{Tkacik+Walczak2011_Review}.
The feasibility of information  as the metric of network performance was recently established by demonstrating that morphogen patterns in early fly development transmit information 
with an accuracy close to the physical limits \cite{Tkacik2008_PNAS,Dubuis2013_PNAS}.
Theoretically, this optimization principle was applied to study information flow in different paradigmatic gene-regulatory scenarios,
including single and multiple target genes regulated by a single input \cite{Tkacik2008_PRE,Tkacik2009_Info,Walczak2010}, feed-forward cross-regulation \cite{Walczak2010},
autoactivation \cite{Tkacik2012}, multistate promotor-switching \cite{Rieckh2014}, 
and time-dependent inputs \cite{Tostevin2009,Tostevin2010,deRonde2010,Mugler2010,Selimkhanov2014}.
These attempts  yielded important insights into optimal strategies with which single cells can reliably respond to external stimuli.
The issue that to-date has not received attention, however, is that many cells or nuclei can respond to a spatially distributed signal collectively,
by exchanging information among themselves, e.g., via diffusion of signaling proteins.
Prominent examples are the specification of emerging body tissues during embryo development \cite{Jaeger2011,Jaeger2012,Rushlow2012,Restrepo2014,Wartlick2009,Meinhardt1982},
the aggregation response in colonies of amoebae \cite{Gregor2010,Kamino2011},
and collective chemosensing in cell colonies and tissues \cite{Sun2012,Sun2013}.
Importantly, spatial aspects not only add another layer of complexity to the distributions of input and output signals of gene-regulatory networks;
they can also markedly alter noise levels---and thus information capacity---through spatial averaging \cite{Erdmann2009,Sokolowski2012,Little2013,Garcia2013,Mugler2014_qbio,Tikhonov2014_qbio}.
A truly predictive theoretical framework  therefore must take into account the spatial setting, 
where multiple regulatory networks can read out the inputs at various locations and exchange information between themselves locally.

Here we extend the existing theory of optimal gene regulatory networks to such a spatial setting.
We construct a spatial-stochastic model of gene regulation that  
takes into account diffusive coupling between neighboring reaction volumes and the relevant sources of noise.
We derive simple expressions for the stationary means and variances of input-activated local protein levels 
as a function of their regulatory parameters and diffusion rate.
From this we compute information transmission in various spatial scenarios and systematically explore
how transmission can be optimized by changing key system parameters.
As a motivating example, we study a model variant representative of the \textit{bicoid-hunchback} system in early fruit fly embryogenesis,
which enables differentiating cells to acquire position-dependent fates
with high reliability and reproducibility in spite of stochasticity in the underlying 
biophysical processes \cite{Houchmandzadeh2002,Gregor2007a,Gregor2007b,Dubuis2013_MSB,Little2013,Petkova2014,Manu2009PlosBiol,DeLachapelle2010}.
Our approach thus specifically optimizes positional information, 
a notion that has been extensively discussed throughout developmental biology \cite{Wolpert1969,Wolpert1994,Wolpert1996,Wolpert2011,Hironaka2012},
but only recently rigorously formalized in mathematical terms, using information theory \cite{Tkacik+Dubuis2015}; the proposed formalism, however, generalizes to other spatially coupled gene regulatory setups.

\pagebreak

We find that spatial coupling can markedly improve information transmission when the dominant source of noise in gene regulation is on the ``input side,'' e.g., due to the random arrival of the regulatory molecules (transcription factors) to their binding sites on the DNA. This is in accordance with the known fact that spatial averaging
can remove super-Poissonian contributions to the noise \cite{Erdmann2009,Sokolowski2012}.
Moreover, with diffusion the maximal information capacity reached upon optimization
 becomes increasingly invariant to the spatial details of the input signal.
Finally, in a defined noise regime that we detail below, 
we observe the emergence of a novel optimal regulatory strategy where diffusion generates graded, and thus informative, 
spatial profiles of regulated genes despite their sharp, threshold-like activation by the regulatory inputs.

\section{Methods}
\label{secMethods}
Here we construct a model in which we can analytically calculate how much information expression levels of diffusible proteins 
encode locally about the position itself, if these proteins are expressed in response to a concentration field of regulatory input molecules.
The outline of our approach is as follows. We {\bf (i)} define a generic spatial-stochastic model of gene regulation that explicitly
accounts for diffusive signaling between adjacent volumes which collectively interpret a spatially distributed input signal (Section~\ref{secMethodsModel}).
Starting from a set of coupled Langevin equations, in Section~\ref{secMethodsMeansVariances} we {\bf (ii)} derive general analytic expressions
for the stationary means and variances of the regulated protein levels,
and provide intuitive insight for the role of diffusion prior to specifying any gene regulatory details.
We then {\bf (iii)} impose realistic  expressions for the gene regulatory function and noise, based on well-established biochemical models (Section~\ref{secMethodsNoiseAndRegF}).
Section~\ref{secMethodsInfo} illustrates how we {\bf (iv)} quantify the encoding of positional information,
using information theory and the stationary moments derived in (ii).
Finally, we vary the parameters of the model to find the optimal regulatory strategies in the spatially coupled setup, which we present in the ``Results'' Section~\ref{secResults}.

\subsection{Spatial-stochastic model of  gene regulation}
\label{secMethodsModel}
\begin{figure}
\begin{center}
  \includegraphics[width=0.5\textwidth]{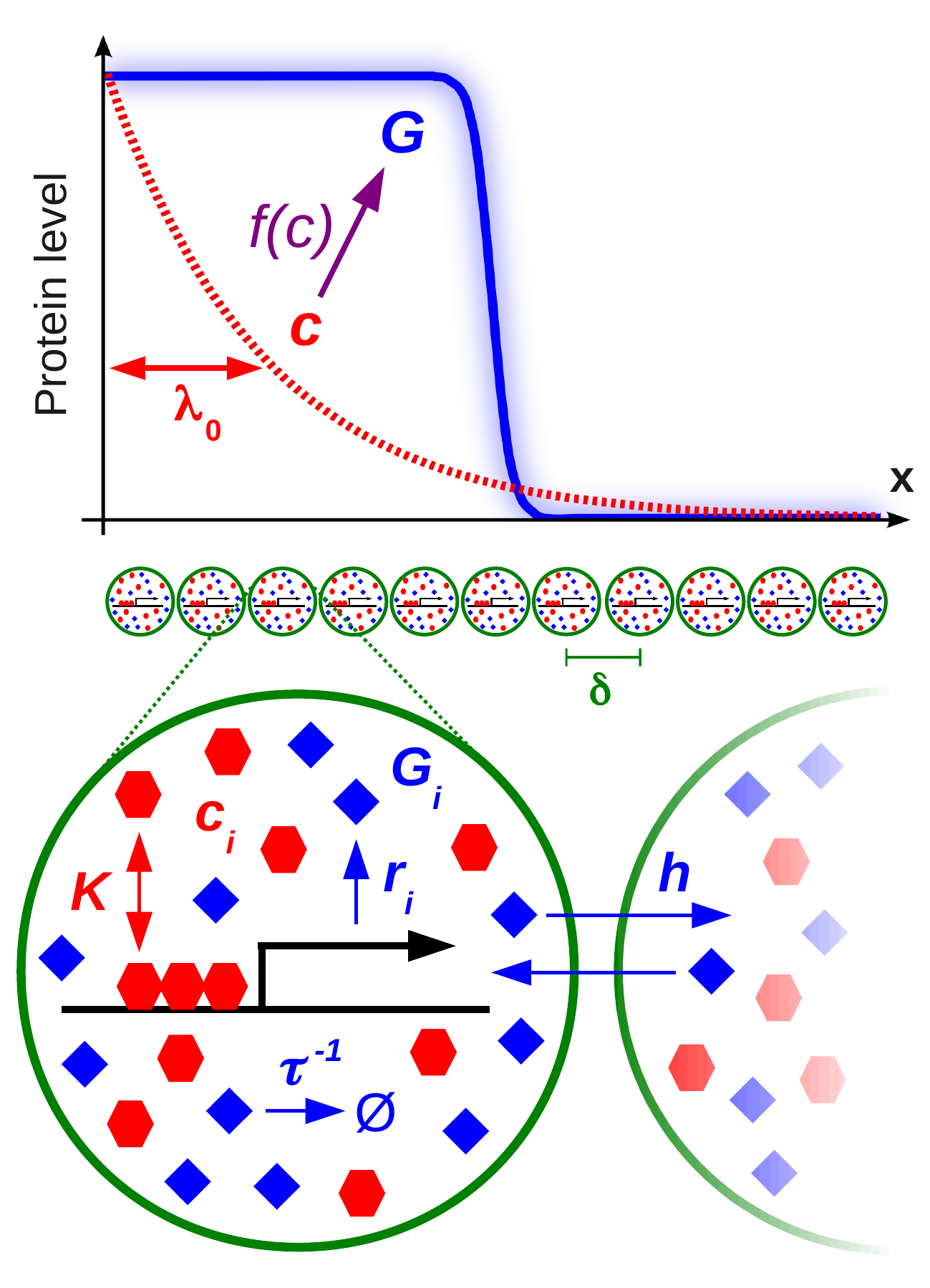}
\end{center}
\caption{{\bf Model schematic.}
The spatial-stochastic model of gene regulation consists of a discrete set
of identical reaction volumes (indicated by $i$) arranged in a regular lattice with lattice spacing $\delta$.
While here for clarity we only draw a 1D lattice,
we also study a 2D model with periodic boundary conditions,
and our framework supports arbitrary coordination numbers.
The volumes collectively sense a spatial signal $c(x)$ which maps the position $x$
to an input concentration $c$ (red), creating a spatial distribution of output proteins $G$ (blue).
Each volume contains an identical promotor activated with probability $f(c_i)$, 
where $f(c)$ is the regulatory function and $c_i$ the activator concentration in volume $i$.
This results in a local production rate $r_i=\rmax f(c_i)$ and local levels $G_i$ of the product protein.
$G$~proteins are degraded with a rate $1/\tau$ and can hop to the neighboring volumes
with a rate $h=D/\delta^2$, where $D$ is their diffusion coefficient.
The regulatory function $f(c)$ is characterized by the activation threshold $K$
and the regulatory cooperativity $H$,
as detailed in Section~\ref{secMethodsNoiseAndRegF}.
}
\label{figModel}
\end{figure}

Figure~\ref{figModel} shows a schematic of the spatially coupled gene regulation model that we study throughout this paper.
We consider a two-dimensional lattice of $N_x \times N_y$ reaction volumes with equal spacing $\ls$ between the volumes.
In this work, we particularly focus on a cylindrical geometry of the lattice,
with periodic boundary conditions along the circumference ($y$-direction), and reflective boundary conditions in axial ($x$-) direction.
While this geometry deliberately mimics the syncytial state of the developing embryo of the fruit fly \Drosophila,
our model could be easily adapted to any arbitrary discrete arrangement of equally spaced reaction volumes.
We will initially consider the special case $N_y=1$ as the ``1D model,'' while referring to the case $N_y>1$ as the ``2D model.''

Each volume contains the promotor of an identical gene controlled by a single transcription factor, whose concentration we denote by $c$. 
The copy number of proteins expressed from this gene of interest will be denoted by $G$, 
while $g=G/\Nmax$ will refer to the copy number normalized by the maximal average expression level $\Nmax$ at full induction.
At the core of our model is the regulatory function $f(c)$ which determines how the activator concentration 
$c$ maps to the effective synthesis rate of $G$ proteins.
We will also refer to $c$ as the ``input,'' and to $G$ and $g$ as the ``output.''
For simplicity we treat transcription and translation as a single production step 
with maximal production rate $\rmax$ reached at full activation.
$G$~molecules are degraded with a constant rate $1/\tau$, where $\tau$ is the average protein lifetime.
In addition, each volume can exchange the product molecules $G$ with its neighboring volumes via diffusion.
Here, we approximate this exchange with an effective ``hopping rate'' $h=D/\ls^2$, where $D$ is the diffusion coefficient of $G$ molecules.

Denoting the position of volume $i$ by $\myvec{x}_i$,
we can write down a stochastic equation for the combined production-degradation-diffusion process in volume $i$ as follows:
\begin{align}
  \partial_t G_i &= \rmax f(c(\myvec{x}_i)) - \frac{1}{\tau}G_i 	\nn\\
		 &\qquad\qquad - \DoverLS\sum_{n\fromNNof{i}} (G_i - G_n) + \sum_k \Gamma_{ik}\eta_k
  \label{eqCoupledLangevin}
\end{align}
Here $G_i$ denotes the number of $G$~proteins in the volume, while the first and second term in the equation describe their production and degradation, respectively.
The third term is a discretized Laplacian accounting for the diffusive exchange of $G$~molecules between volume $i$ and its nearest-neighbor volumes $n\fromNNof{i}$.
Function $c(\myvec{x})$ maps positions $\myvec{x}_i$ to local transcription factor concentrations $c_i$.
In this work we consider the case in which $c(\myvec{x})$ is invariant in $y$-direction, i.e. $c_i \equiv c(\myvec{x}_i) = c(x(i))$;
we specify the functional form of $c(x)$ at the end of this section.
The noise term $\sum_k \Gamma_{ik}\eta_k$ comprises all noise sources $\eta_k$ that act on the protein level $G_i$,
including those outside of volume $i$.
We assume that all fluctuations are well represented by non-multiplicative, zero-mean Gaussian white noise, and compute the relevant noise powers $\Gamma_{ik}$ below.

\subsection{Means and variances with spatial coupling}
\label{secMethodsMeansVariances}
The set of coupled Langevin equations [Eq.~(\ref{eqCoupledLangevin})] is a special case of 
what is known as the ``N-unit generalized Langevin model'' \cite{Hasegawa2006a,Hasegawa2007} in other contexts.
To compute the steady-state means and variances we employ a technique based on It\={o} calculus
which allows to convert the set of coupled stochastic equations for the variables $G_i$ 
into a set of ODEs for their moments that hold exactly.
This can be performed for an arbitrary discrete coupling network, 
and even for the case of multiplicative noise (not considered here) \cite{Rodriguez1996,Hasegawa2006a,Hasegawa2007}.

We can write down the equations of motion for the mean of the normalized local output $g_i=G_i/\Nmax$ 
and the (co)variances of its fluctuations $\dg_i \equiv g_i - \la g_i \ra$ as follows \cite{Rodriguez1996,Hasegawa2006a,Hasegawa2007}:
\begin{align}
\partial_t \la g_i \ra &= \la F(g_i,c_i) \ra - \hoprate \sum_{n_i\fromNNof{i}} \big\la g_i - g_{n_i} \big\ra		\label{eqODEmean}\\
\partial_t \la \dg_i \dg_j \ra 	&= \la \dg_i F(g_j,c_j) \ra + \la \dg_j F(g_i,c_i) \ra 							\nn\\
				  &\quad + \hoprate \left\la \dg_i \sum_{n_j\fromNNof{j}} (g_{n_j} - g_j) \right\ra			\nn\\
				  &\quad + \hoprate \left\la \dg_j \sum_{n_i\fromNNof{i}} (g_{n_i} - g_i) \right\ra	 		\nn\\
				  &\quad + \sum_k \la \Gamma_{ik} \Gamma_{jk} \ra					\label{eqODEvar}
\end{align}
Here the function $F(g_i,c_i) = \frac{1}{\tau} \left( f(c_i) - g_i \right)$ groups the production and degradation terms
and depends on the regulatory function $f(c)$.
The $n_i$- and $n_j$-sums run over the nearest-neighbor volumes of $i$ and $j$, respectively.
Note that in the $k$-sum running over all noise sources in the system usually most terms will be zero.

The hitherto arbitrary noise powers $\Gamma_{ik}$ can be written out more explicitly by listing all noise sources 
that affect volume $i$ in the following way:
\begin{align}
 \sum_k \Gamma_{ik}\eta_k =& \GNoise_i(c_i,\myvec{g})\gamma_i		\nn\\
			   & - \sum_{n\fromNNof{i}} \DNoise_{(i\rightarrow n)} \xi_{(i\rightarrow n)} 
			     + \sum_{n\fromNNof{i}} \DNoise_{(n\rightarrow i)} \xi_{(n\rightarrow i)}
\label{eqNoiseComposition}
\end{align}
Herein 
$\gamma_i$ is the noise in the combined process of random regulation, production and degradation,
and $\xi_{(i\rightarrow j)}$ the noise caused by random hopping from volume $i$ to $j$.
A crucial step in specifying the expression above is to write the noise powers of the incoming and outgoing hopping processes
with different signs, because each hopping event causing a negative fluctuation in the volume
of origin always causes a positive fluctuation in the volume of arrival.
The hopping thus is a birth process when seen from the volume of arrival, and a death process when seen from the volume of origin.
This intuitive argument can be derived more rigorously from the Langevin equation 
following the method portrayed by Gillespie \cite{Gillespie2000},
and is quantitatively confirmed by stochastic simulations, employing (e.g.) the next-subvolume method \cite{Hattne2005,Elf2004}.

Using Eq.~(\ref{eqNoiseComposition}) we can rewrite the noise term $\sum_k \la \Gamma_{ik} \Gamma_{jk} \ra$ in Eq.~(\ref{eqODEvar})
for the variances (case $i=j$) and nearest-neighbor covariances (case $i\fromNNof{j}$) as follows:
\begin{align}
 \sum_k \left\la\Gamma_{ik}\Gamma_{jk}\right\ra	&= \la \GNoise_i^2(c,\myvec{g}) \ra
						   + \sum_{n\fromNNof{i}} \la\DNoise_{(i\rightarrow n)}^2 + \DNoise^2_{(n\rightarrow i)}\ra 	\nn\\
						&\equiv \Noise^2_{ii} \qquad\qquad\qquad\qquad \text{for } i=j							\\
 \sum_k \left\la\Gamma_{ik}\Gamma_{jk}\right\ra	&=  \underset{\text{hopping } (i\rightarrow j)}{ \underbrace{\left\la\Gamma_{i\kappa}\Gamma_{j\kappa}\right\ra} }
						  + \underset{\text{hopping } (j\rightarrow i)}{ \underbrace{\left\la\Gamma_{i\kappa'}\Gamma_{j\kappa'}\right\ra} } 	\nn\\
						&= -\la\DNoise^2_{(i\rightarrow j)}\ra - \la\DNoise^2_{(j\rightarrow i)}\ra 						\nn\\
						&\equiv \Noise^2_{ij} \qquad\qquad\qquad\qquad \text{for } i\fromNNof{j}
\end{align}

To obtain the stationary means and covariances of the protein number $g_i$,
we solved Eqs.~(\ref{eqODEmean}) and (\ref{eqODEvar}) in steady state
by setting their left sides to zero ($\partial_t \la .. \ra = 0$).
After substituting $g_i = \la g_i \ra + \dg_i$,
we regroup covariances $\la \dg_i\dg_j \ra$ and means $\la g_i \ra$ 
and  relate them to the respective quantities in the neighboring volumes
(see Appendix~\ref{secAppendixMeansVariances}).
In this context, we introduce the pragmatic assumption 
that only concentrations in volumes that are nearest neighbors on the lattice 
have significant correlations, i.e., we enforce 
\begin{align}
  \la \dg_i \dg_j \ra = 0 \qquad\quad \text{for } j\notfromNNof{i}
\end{align}
While this ``short correlations assumption'' markedly facilitates both analytical progress and computation speed,
and yields formulas that are intuitive to interpret,
our results only change marginally when the assumption is released (see Section~\ref{secSpatialCouplingModelComparison}).
For brevity, in the following we write
      \begin{align}
      \mean_i		\equiv \la g_i \ra,		\qquad
      \sigma^2_i	\equiv \la \dg_i^2 \ra,		\qquad
      \cc_{ij}		\equiv \la \dg_i\dg_j \ra
      \end{align}
for the steady-state means, variances and covariances, respectively.
Treating the cases $i=j$ and $i\fromNNof{j}$ in Eq.~(\ref{eqODEvar}) separately,
we arrive at the following set of coupled equations for these quantities (cf. Appendix~\ref{secAppendixMeansVariances}):
\begin{align}
\mean_i		&= \frac{\tau}{1+2d\Delta} \left( r_i + \hoprate\sum_{n_i\fromNNof{i}} \mean_{n_i} \right)	\nn\\
		&\equiv T r_i + \Lambda^2 \sum_{n_i\fromNNof{i}} \mean_{n_i}					\label{eqMeanSS}\\
\sigma^2_i 	&= \Lambda^2 \sum_{n_i\fromNNof{i}} \cc_{in_i} + \frac{T}{2} \NNoise_{ii}^2 			\label{eqVarSS}\\
\cc_{ij}   	&= \frac{\Lambda^2}{2} \left( \sigma^2_i + \sigma^2_j \right) + \frac{T}{2} \NNoise_{ij}^2	\label{eqCorrSS}
\end{align}
Here we abbreviate normalized production rate $r_i = f(c_i) / \tau$;
$d$ is the lattice dimension, or half of the coordination number of a reaction volume.
We  introduce  the dimensionless diffusion constant $\Delta\equiv D/D_0$, 
which is equal to the diffusion constant measured in the ``natural'' unit $D_0=\delta^2/\tau$; note that $\Delta=h\tau$.
We will also refer to $\Delta$ as the ``spatial coupling.''

The above equations define two ``effective parameters'': 
 {\bf (i)}, an ``effective residence time'' $T=\tau/(1+2d\Delta)$, reflecting the fact that with diffusion proteins are taken out from
the reaction volume faster than through regular degradation in the absence of diffusion;
and {\bf (ii)}, the (dimensionless) ``mixing parameter'' $\Lambda = \sqrt[+]{\hoprate T} = \sqrt[+]{D T} / \ls = \sqrt{\Delta / (1+2d\Delta)}$,
which equals the distance travelled via diffusion during the time $T$, measured in units of the lattice spacing $\delta$.
These quantities capture the essential effect of diffusive coupling between the stochastic processes in neighboring reaction volumes:
With increasing coupling ($\Delta>0 \Rightarrow \Lambda^2 > 0$), the mean output level $\mean_i$ is increasingly 
set by the mean levels in the neighboring volumes $\mean_{n_i}$, 
whereas the contribution of the local production to the apparent copy number in the volume is reduced ($T < \tau$).
An analogous interpretation holds for $\sigma_i^2$ if $\frac{1}{2} \NNoise_{ii}^2$ is interpreted as a ``noise production'' term.

We verified that Eqs.~(\ref{eqMeanSS}), (\ref{eqVarSS}) and (\ref{eqCorrSS}) correctly
reproduce the known result that all super-Poissonian noise is attenuated in the ``Poissonian limit,''
meaning that (in non-normalized units) the variance becomes equal to the mean 
for inifinitely strong coupling $\Delta \rightarrow \infty$ \cite{Erdmann2009,Sokolowski2012}.
This holds both in 1D and 2D, and with or without short-correlations assumption.
 
Equations~(\ref{eqMeanSS}), (\ref{eqVarSS}) and (\ref{eqCorrSS}) can be solved for the variables $\mean_i$, $\sigma_i^2$ and $\cc_{ij}$
after specifying the detailed forms of the regulatory function and noise strength and imposing meaningful boundary conditions.
If the regulatory function and noise do not contain terms nonlinear in the variables of interest,
Eqs.~(\ref{eqMeanSS}), (\ref{eqVarSS}) and (\ref{eqCorrSS}) form a set of coupled linear equations that can be solved by simple matrix inversion.
Also notice that we may use Eq.~(\ref{eqCorrSS}) to simplify Eq.~(\ref{eqVarSS}) further 
if we are only interested in solving for the variances $\sigma^2_i$ (as in this work);
we show the respective formulas for the 1D and 2D model in Appendix~\ref{secAppendixVarianceSolutions}.

\subsection{Regulatory function and gene expression noise}
\label{secMethodsNoiseAndRegF}
To describe the probability of the promotor being in its active state given that the concentration of the transcription factor is $c$, 
we choose a simple Hill-type regulation model:
\begin{align}
 f(c) = \frac{c^H}{c^H + K^H}, 
 \label{eqRegFunction}
\end{align}
where $H$ is the Hill coefficient, quantifying the cooperativity of the activation process,
and $K$ the activation threshold, i.e. the concentration at which half-activation occurs, $f(K)=1/2$. 

To complete our analytical model we need to accurately specify the noise powers $\la\GNoise^2_{i}\ra$ and $\la\DNoise^2_{(i\rightarrow j)}\ra$ 
appearing in the noise terms $\NNoise_{ii}^2$ and $\NNoise_{ij}^2$.
In steady state, all of these noise powers only depend on the mean levels of the transcription factor ($\lbrace c_i \rbrace$) and product ($\lbrace \mean_i \rbrace$).

Let us first specify the power of the regulation/production/degradation noise, $\la\GNoise^2_{i}\ra$,
which has two contributions:
super-Poissonian ``input noise'' $\la\GNoise^2_{TF,i}\ra$ originating from random arrival of regulatory transcription factors,
and Poissonian ``output noise'' $\la\GNoise^2_{PD,i}\ra$ from the production and degradation of the gene product.
The input noise is a function of the transcription factor concentration and is well-approximated by \cite{Tkacik2008,Tkacik2009_Info,Tkacik2009_Diff}
\begin{align}
 \la\GNoise^2_{TF,i}\ra = \left[\rmax^2 \left( \frac{\partial f(c)}{\partial c} \right)^2 \frac{2c}{D_c l_c} \right]_{c=c_i} \label{inoise}
\end{align}
where $D_c$ is the diffusion coefficient of the transcription factors, $l_c$ the typical size
of a transcription factor binding site and $c_i$ the local activator concentration.
The above formula describes fluctuations of $G_i$ in absolute units;
to obtain the noise power $\la\hat\GNoise_{TF,i}\ra$ of fluctuations in the normalized output $g_i$,
$\la\GNoise^2_{TF,i}\ra$ has to be normalized as well by $\Nmax^2=(\rmax\tau)^2$:
\begin{align}
 \la\hat\GNoise^2_{TF,i}\ra &= \frac{\la\GNoise^2_{TF,i}\ra}{\Nmax^2}
			     = \frac{1}{\Nmax\tau} \left[\left( \frac{\partial f(c)}{\partial c} \right)^2 \frac{2c\Nmax}{D_c l_c \tau} \right]_{c=c_i}	\nn\\
			    &= \frac{1}{\Nmax\tau} \left[\left( \frac{\partial f(c)}{\partial c} \right)^2 2c c_0 \right]_{c=c_i}
\end{align}
In the last step, $c_0 \equiv \Nmax / (D_c l_c \tau)$ defines a typical concentration scale for our problem;
in the following we will measure concentrations in units of $c_0$.

The output noise in absolute units can be written as the sum of production and degradation rate,
as for a regular birth/death process:
\begin{align}
 \la\GNoise^2_{PD,i}\ra = \rmax f(c_i) + \frac{1}{\tau} \la G_i \ra
\end{align}
Note that in general $\la G_i \ra \neq \rmax f(c_i)\tau$ due to the diffusive coupling.
Also note that bursty production (not considered in this work) could be easily incorporated into the model
via a prefactor $\phi$ (Fano factor) in the production term: $\rmax f(c_i) \rightarrow \phi\rmax f(c_i)$.
In normalized units the production/degradation noise reads:
\begin{align}
 \la\hat\GNoise^2_{PD,i}\ra = \frac{\la\GNoise^2_{PD,i}\ra}{\Nmax^2} = \frac{1}{\Nmax \tau} \left( f(c_i) + \mean_i \right)
\end{align}

In our model, diffusive hopping of a molecule is identical with its degradation in the volume of origin
and simultaneous production of a molecule in the volume of arrival.
The corresponding noise powers $\DNoise^2_{(i\rightarrow j)}$ thus are given by
the rate of particle loss through diffusive hopping from the volume of origin,
which in steady state is given by the mean copy number in that volume times the hopping rate,
and can be normalized in the same way as the other noise powers:
\begin{align}
 \la\DNoise^2_{(i\rightarrow j)}\ra = \frac{D}{\ls^2} \la G_i \ra, \quad\; 
 \la\hat\DNoise^2_{(i\rightarrow j)}\ra = \frac{\la\DNoise^2_{(i\rightarrow j)}\ra}{\Nmax^2} = \frac{\Delta}{\Nmax\tau} \mean_i	\nn\\
\end{align}

Adding up all contributions, we obtain for the (normalized) combined noise powers $\NNoise^2_{ij}$ and $\NNoise^2_{ii}$:
\begin{align}
\NNoise^2_{ij} &= -\la\DNoise^2_{(i\rightarrow j)}\ra -\la\DNoise^2_{(j\rightarrow i)}\ra = -\frac{\Delta}{\Nmax\tau} \left( \mean_i + \mean_j \right)	\label{eqNormNoisePowerCov}\\
\NNoise^2_{ii} &= \la\GNoise^2_{PD,i}\ra + \la\GNoise^2_{TF,i}\ra + \sum_{n\fromNNof{i}} \la\DNoise_{(i\rightarrow n)}^2 + \DNoise^2_{(n\rightarrow i)}\ra	\nn\\
	       &= \frac{1}{\Nmax\tau} \vast( f(c_i) + \mean_i + \left[\left( \frac{\partial f(c)}{\partial c} \right)^2 2c c_0 \right]_{c=c_i} 		\nn\\
	       &\hspace{3.35cm} + \Delta\sum_{n\fromNNof{i}}\left( \mean_i + \mean_n \right) \vast)
		\label{eqNormNoisePowerVar}
\end{align}

\subsection{Quantifying positional information}
\label{secMethodsInfo}
Building on our previous work \cite{Tkacik2008_PRE,Tkacik2008_PNAS,Tkacik2009_Info,Walczak2010,Tkacik2012}, 
we quantify the amount of information that the noisy output signal $g(x)$ carries about the position $x$ 
using the mutual information $I(x;g)$ between the input and output \cite{Shannon1948,Kolmogorov1965},
a central quantity of Shannon's information theory \cite{Shannon1948}:
\begin{align}
I(x;g) = \int dx \,P_x(x) \int dg \, P(g|x) \log_2\left[ \frac{P(g|x)}{P_g(g)} \right]
\label{eqMIdefinition}
\end{align}
In the information-theoretic picture introduced here,
the information channel is defined by a set of sampling units (cells or nuclei at a particular position) that implement the same gene regulatory network.
These sampling units read out the input signal $c$ with a distribution that is jointly determined by the shape of the concentration field, $c(x)$,
and the spatial distribution of the sampling units, $P_x(x)$.
In response to these inputs, the regulatory networks locally express the output gene $g$ at the corresponding level,
producing an output distribution $P_g(g)$ across the ensemble of sampling units.

The mutual information can be rewritten as a difference between the entropy of the output distribution $S[P_g(g)] = -\int dg \, P_g(g) \log_2 P_g(g)$
and the average entropy of the conditional distribution of $g$ at fixed position $x$:
\begin{align}
I(x;g) = S[P_g(g)] - \int dx \, P_x(x) S[P(g|x)]
\label{eqMIdiff}
\end{align}
When only the local distribution of outputs $P(g|x)$ is known, $P_g(g)$ can be straightforwardly constructed from $P(g|x)$ and $P_x(x)$:
\begin{align}
P_g(g) = \int dx \, P_x(x) P(g|x)
\label{eqPgExpand}
\end{align}
In our discrete model we assume uniform spacing of the sampling positions $x_i \in [0,N_x\delta]$,
i.e. $P_x(x_i) = 1/N_x$. Equations~(\ref{eqMIdiff}) and (\ref{eqPgExpand}) then respectively become:
\begin{align}
I(x;g)	&= -\int dg \, P_g(g) \log_2 P_g(g)  							\nn\\
        &\qquad +  \frac{1}{N_x} \sum_{i=1}^{N_x} \int dg \, P(g|x_i) \log_2 P(g|x_i)		\label{eqMIgSum}	\\
P_g(g) 	&=  \frac{1}{N_x} \sum_{i=1}^{N_x} P(g|x_i)						\label{eqMIgSumPg}
\end{align}       
$I(x;g)$  is therefore fully characterized by the conditional output distribution $P(g|x_i)$.
Since here we only consider non-multiplicative white noise, $P(g|x_i)$ is Gaussian:
\begin{align}
P(g|x_i) = \frac{1}{\sqrt{2\pi\sigma_g^2(x_i)}} \exp\left(-\frac{\left( g - \bar g(x_i) \right)^2}{2\sigma_g^2(x_i)} \right)
\end{align}
The information $I(x;g)$  is therefore fully determined by the (local) means and variances of the conditional distributions $P(g|x_i)$.

In our previous work \cite{Tkacik2008_PRE,Tkacik2008_PNAS,Tkacik2009_Info,Walczak2010,Tkacik2012}, 
we have computed the maximum achievable information transmission (channel capacity) by optimizing over the distribution of input signals, 
relying on the ``small noise approximation'' to make the problem analytically tractable.
Here, in contrast, the spatial setting has led us to choose a uniform distribution of sampling points, $P_x(x)$.
While this relieves us of the need to use the small noise approximation,
to fully specify the problem
we still need to select the function $c(x)$ that maps the input of the regulatory network to spatial positions.
Motivated by the developmental example of the fruit fly, here we choose an exponential function \cite{Houchmandzadeh2002,Gregor2007a}:
\begin{align}
 c(x) = \cmax e^{-x/\left( \lambda \gamma_0 \right)},
\end{align}
where $\cmax$ controls the maximum achievable input, 
$\gamma_0 \equiv L/5 \equiv N_x\delta/5$ is the decay length of a typical exponential morphogen gradient 
relative to the system size $L$ \cite{Houchmandzadeh2002,Gregor2007a,Bollenbach2008},
and $\lambda$ a variable dimensionless scaling factor (the precise value of $\gamma_0$ is insignificant; we could opt for any convenient length unit). 
This choice generates a family of input profiles parametrized by $\cmax$ and $\lambda$,
and we will explore various settings for these parameters to maximize the positional information encoded by the expression level $g$.

We can now assemble the different components of the model.
By inserting the regulation and noise terms [Eqs.~(\ref{eqRegFunction}), (\ref{eqNormNoisePowerCov}) and (\ref{eqNormNoisePowerVar})] 
into the steady-state solutions of the coupled stochastic equations [Eqs.~(\ref{eqMeanSS}), (\ref{eqVarSS}), (\ref{eqCorrSS})]
we obtain two coupled linear equation systems for mean expression levels $\mean_i$ and (co)variances $\cc_{ij}$ and $\sigma^2_i$ which we can solve given the selected boundary conditions.
Using Eq.~(\ref{eqMIgSum}) we can then compute the mutual information $I(x;g)$.
When computing the stochastic moments, 
we can vary the parameters of the input function, the regulation and the spatial coupling,
and thus compute $I(x;g)$ as a function of its key determinants.
The baseline set of parameters that we used for numerical computation
is presented in Appendix~\ref{secAppendixParameters}.

\section{Results}
\label{secResults}
In order to elucidate how spatial coupling alters the capacity of encoding positional information
in the downstream gene $g$, we optimized the mutual information $I(x;g)$ over the parameters of our model: 
the dissociation constant, $K$, and the Hill coefficient of regulation, $H$; 
the (normalized) diffusion constant of the gene product, $\Delta$; and the length scale of the input gradient, $\lambda$. 
As is known from previous work, the key parameter that qualitatively determines the shape of the optimal solutions is the ratio of the output to input noise, 
which is set by a dimensionless maximal input concentration, $C= \cmax/c_0$ \cite{Tkacik2009_Info}; 
in general, $C\gg 1$ is the regime of dominant output noise, while $C\ll 1$ is the regime of dominant input noise, 
as defined in Section~\ref{secMethodsNoiseAndRegF}. 
We thus explored the optimal solutions as a function of $C$ in what follows.

For computational efficiency we initially studied the 1D model with short-correlations assumption (SCA), and later
compared to the 2D models with and without SCA, finding only minor differences (see Section~\ref{secSpatialCouplingModelComparison}).

\begin{figure*}[ht]
\begin{center}
   \includegraphics[width=\textwidth]{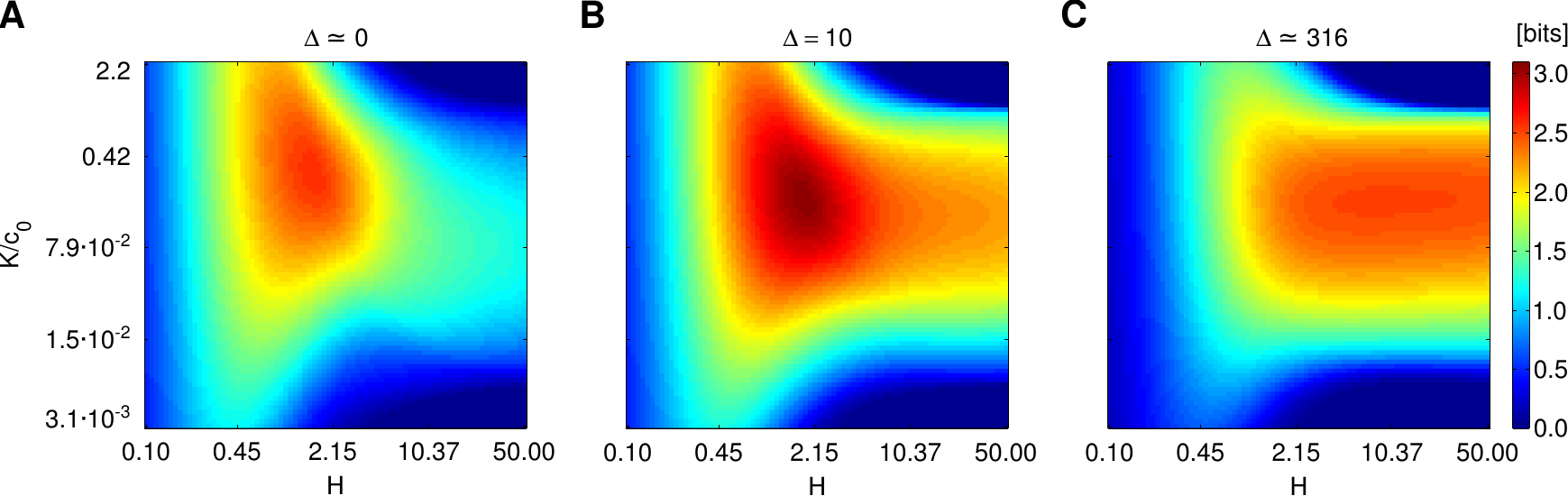}
\end{center}
\caption{{\bf Information planes for different values of the spatial coupling, $\Delta$.}
	  Shown is the mutual information $I(x;g)$ as a function of the Hill coefficient $H$ and the activation threshold $K$,
	  {\bf (A)} with negligible spatial coupling, $\Delta \simeq 0$;
	  {\bf (B)} with intermediate spatial coupling, $\Delta = 10$; and
	  {\bf (C)} with very strong spatial coupling, $\Delta \simeq 316$.
	  The data shown is for the 1D model with short-correlations assumption.
	  }
\label{figEffectOfCoupling}
\end{figure*}

\subsection{Spatial coupling can enhance information transmission}
\label{secSpatialCouplingComparison}
We first assessed how introducing diffusive coupling changes the information capacity of the system
in the  space of regulatory parameters $K$ and $H$.
To establish the baseline for comparison, we started with the case without spatial coupling, $\Delta\simeq0$,
fixing $C=1$ and input gradient length $\lambda=1$. 
Figure~\ref{figEffectOfCoupling}A shows the ``information plane'' for this case,
i.e., the mutual information $I(x;g)$ as a function of the Hill coefficient $H$ and the activation threshold $K$. 
Consistently with our previous studies \cite{Tkacik2008_PRE,Tkacik2009_Info},
the information plane displays a clear optimal choice of $H$ and $K$ that maximizes information transmission. 
The optimum results from a nontrivial compromise between evenly spreading the whole dynamic range of the output signal along the $x$-axis,
which ideally requires low Hill coefficients and half-activation at the system center ($x=L/2, K=c(L/2)$),
and a system-wide minimization of the noise;
the latter generally favors activation at high input concentrations (to avoid large fluctuations at low $c$) and thus higher $K$. 

Figures \ref{figEffectOfCoupling}B and \ref{figEffectOfCoupling}C show the same information plane, 
but now at increasing values for the spatial coupling $\Delta$. 
The first observation is that with nonzero $\Delta$ information can be increased relative to the baseline at the optimal choice of $H$ and $K$, 
but that the information drops again  when the diffusive coupling is too strong, 
indicating that there exists a nontrivial optimal value for $\Delta$ that maximizes information. 
The second observation relates to the concomitant change in the optimal parameters $\{K^*, H^*\}$ as the diffusive coupling is increased. 
In particular, increasing the diffusion constant shifts the optimum towards higher $H$ and lower $K$. 
The third observation is that strong diffusion also increases the capacity away from the optimum in the regime of high $H$, 
creating a flat plateau where the precise value of $H$ is not crucial for attaining high information transmission.

All these observations can be understood intuitively:
The increase in capacity for nonzero diffusion is due to the trade-off between the ability of diffusive spatial averaging to reduce noise (thus increasing the transmission), 
and smoothing out the response profile (thus decreasing the transmission) \cite{Erdmann2009,Sokolowski2012}.
The shift in optimal regulatory parameters is a direct consequence of these two effects: the detrimental effect of smoothing out the response profile 
can be partially compensated for by choosing  higher optimal Hill coefficients, 
while noise reduction allows the optimal $K$ to move towards lower concentrations to better utilize the dynamic range. 
The plateau in information at high diffusion results from the ability of the diffusion to flatten sharp, 
nearly step-like gene activation profiles at high $H$ (limited to at most 1 bit of positional information for $H\rightarrow\infty$ and $\Delta\rightarrow 0$) 
into smooth spatial gradients that can convey more positional information.

\subsection{Spatial coupling is most beneficial when input noise is dominant}
\label{secSpatialCouplingSystematic}
\begin{figure*}
\begin{center}
  \includegraphics[width=\linewidth]{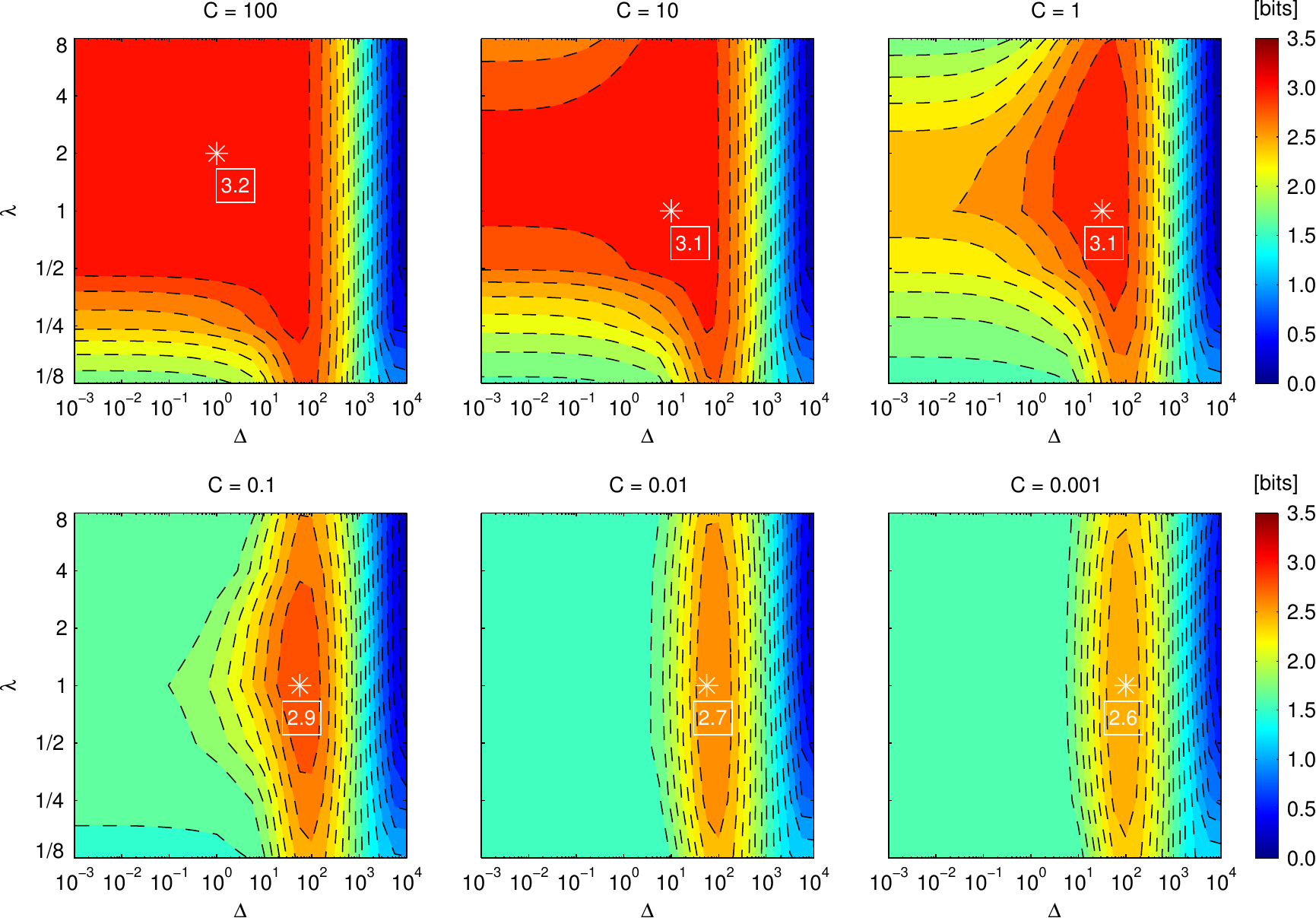}
\end{center}
\caption{{\bf Information transmission as a function of the spatial coupling $\Delta$ and 
	  the input gradient length $\lambda$ for various values of maximal input concentration, $C$.}
	  For comparison, the same scale is used for all contour maps.
	  White stars mark the overall maximum of $I(x;g)$ in each respective panel;
	  white boxes show the corresponding optimal amount of transmitted information (in bits).
	  Each datapoint was obtained by optimizing $I(x;g)$ over the regulatory parameters $H$ and $K$
	  for the given set of $C$, $\Delta$ and $\lambda$.
	  The data shown is for the 1D model with short-correlations assumption.
	  }
\label{figOptimumVsDAndLF}
\end{figure*}
After understanding the optimal behavior of information as a function of regulatory parameters $H,K$ at fixed $C=1$, $\lambda=1$, and $\Delta$, 
we systematically varied the diffusive coupling $\Delta$ and the input gradient length scale $\lambda$ for several different values of $C$. 
For each choice of $(\Delta,\lambda)$ parameters, we separately mapped out the information as a function of $H$ and $K$ to find the optimal values $(H^*,K^*)$ 
and the corresponding maximal information transmission.

Figure~\ref{figOptimumVsDAndLF} summarizes this exploration and shows the information capacity as a function of $\Delta$ and $\lambda$ 
for a series of 6 different $C$ values. White stars mark the maximal capacity for the given $C$,
i.e., the capacity for jointly optimal choice of $H$, $K$, $\Delta$ and $\lambda$.

The panel for $C=1$ demonstrates two beneficial effects of diffusion on information capacity:
increasing $\Delta$ from very low values to $\Delta\lesssim100$ increases $I(x;g)$ and simultaneously
enlarges the range of $\lambda$ over which high capacity can be reached.
This effect is even more pronounced when $C<1$: 
at very low $C$ the beneficial values for the diffusion constant $\Delta$ are strongly constrained to a narrow interval, 
and the resulting capacity gain at optimal $\Delta$ with respect to low $\Delta$ increases markedly.
In contrast, when $C$ is high, diffusive coupling does not convey a large benefit in information transmission. 
This is due to the fact that diffusion can only remove super-Poissonian parts of the noise \cite{Erdmann2009, Sokolowski2012}. 
In our model, super-Poissoninan noise is the input noise, which plays a minor role for $C\gg 1$, 
thereby limiting the potential role of diffusion. 
We note that super-Poissonian fluctuations could also occur on the output side 
(e.g., when gene expression is bursty and multiple protein copies are produced from each mRNA), 
in which case diffusive coupling could be beneficial even for $C\gg 1$.
As expected, for all $C$, the information capacity drops to zero at very high $\Delta$ independently of all other parameters, 
because in that limit all output profiles become flat and thus uninformative.

\begin{figure}
\begin{center}
  \includegraphics[width=\linewidth]{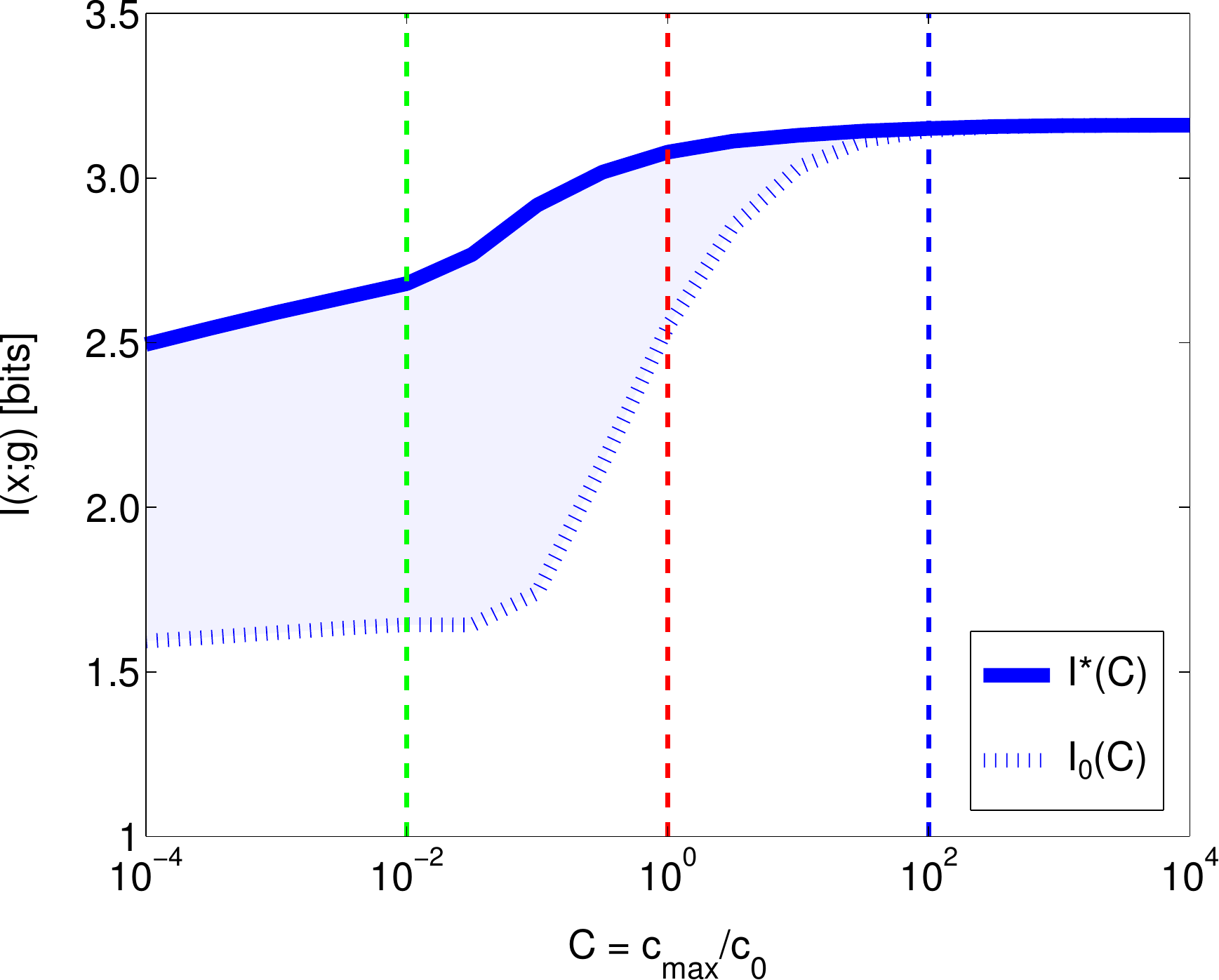}
\end{center}
\caption{{\bf Information capacity for varying maximal input concentration, $C$.}
	 Shown is the optimized information capacity as a function of $C=\cmax/c_0$
	 with optimal diffusive coupling ($I^*(C)$, solid blue line)
	 and without diffusive coupling ($I_{0}(C)$, dashed blue line).
	 At small $C$, non-Poissonian input noise is dominant, while Poissonian output noise dominates for large $C$.
	 The blue-shaded area depicts the maximal gain in information capacity from diffusive coupling:
	 spatial coupling enhances information capacity most efficiently when input noise dominates.
	 The dashed vertical lines mark the $C$ values corresponding to the three cases displayed in Fig.~\ref{figOptimalProfiles}. 
}
\label{figImaxVsC}
\end{figure}

We systematically explored how diffusion affects the optimal information transmission as a function of $C$ in Fig.~\ref{figImaxVsC}, 
by comparing the optimally coupled spatial model to a model where diffusion is set to zero.
Because the information transmission is largely invariant to $\lambda$ and for most values of $C$ peaks in a very broad plateau at $\lambda=1$, 
we fixed $\lambda=1$ for this comparison, while optimizing over all other parameters (separately for the spatially coupled system and the system with $\Delta\simeq 0$). 
Clearly, in the low-$C$ regime optimal diffusive coupling can enhance information capacity by more than a bit,
while for $C\gtrsim100$ the noise composition is strongly dominated by Poissonian output noise such that diffusion cannot 
further improve information throughput.

Instead of varying $C$ by changing the maximal input concentration $\cmax$,
we also varied $C$ via $\Nmax$, thus altering output noise at constant levels of input noise.
In that case we observe the same behavior, i.e. significant capacity enhancement by diffusion in the low-$C$ regime,
with the small difference that now the capacity increases with decreasing $C \sim 1/\Nmax$,
as presented in more detail in Appendix~\ref{secAppendixVaryNmax}.

\subsection{Spatial coupling enables a novel regulatory strategy at high input noise}
\label{secSpatialCouplingEffectOnProfiles}

\begin{figure}
\begin{center}
  \includegraphics[width=\linewidth]{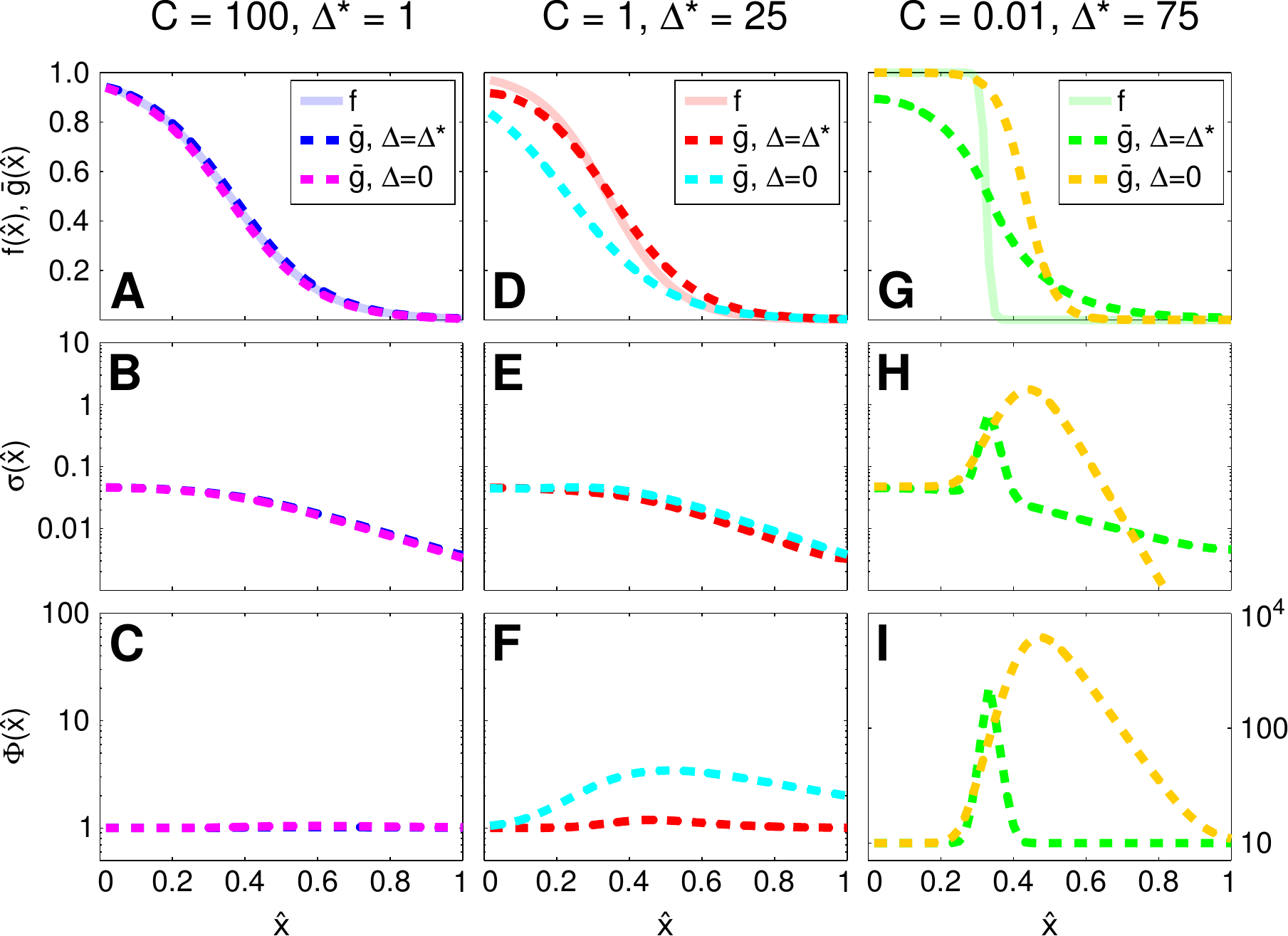}
\end{center}
\caption{{\bf Comparison of optimal output profiles for different values of the maximal input concentration, $C$. } {\bf (A-C)} Dominant output noise, $C=100$, blue lines. 
{\bf (D-F)} Balanced noise, $C=1$, red lines. {\bf (G-I)} Dominant input noise, $C=0.01$, green lines. 
For every value of $C$,  the normalized mean output profile $\mean(\hat x)$ (dashed lines) and the optimal activation profile $f(\hat x)$ (faint solid lines) are shown in the top row; 
the noise in gene expression is shown as the standard deviation $\sigma(\hat x)$ in the middle row; 
and the Fano factor $\Phi(\hat x)=\Nmax \sigma^2(\hat x)/\mean(\hat x)$, is shown in the bottom row. 
All profiles are shown as a function of the normalized position, $\hat x = x/L$. 
The corresponding quantities for the optimal spatially uncoupled ($\Delta \simeq 0$) system are shown in complementary colors (see plot legends). 
}
\label{figOptimalProfiles}
\end{figure}

Spatial coupling affects both the noise level of the output gene as well as the shape of its output profile. We wondered how these two effects combine to affect the information transmission. 
Figure~\ref{figOptimalProfiles} shows the optimal mean output profile $\mean(\hat x)$,
the underlying activation profile $f(\hat x)$,
and two measures of the noise in gene expression, the standard deviation $\sigma^2(\hat x)$
and the Fano factor $\Phi(\hat x)=\Nmax \hat\sigma^2(\hat x)/\mean(\hat x)$, 
as a function of position $\hat x\equiv x/L \in [0,1]$. 
We plot these quantities for three values of $C$, representative of the input-noise-dominated regime ($C=0.01$),
the regime in which input and output noise are approximately balanced ($C=1$),
and the output-noise-dominated regime ($C=100$).
In each plot we directly compare the system in which the diffusion constant is optimized along with all other parameters, 
vs. the system that is optimal at zero spatial coupling. 

For $C=100$, input noise levels are so low that the total noise is dominated by purely Poissonian output noise throughout the whole system ($\Phi(\hat x)\simeq 1$).
Therefore, strong spatial coupling would fail to decrease the noise further and would merely fade out the mean output profile.
Consequently, the optimal diffusion constant $\Delta^*=1$ is low, and the spatially coupled system only differs marginally from the uncoupled system.

For $C=1$ in the uncoupled system, input noise becomes more prominent as a fraction of the total noise, 
such that $\Phi(\hat x)>1$ for $\hat x\gtrsim0.3$ (cyan colored line in Fig.~\ref{figOptimalProfiles}F);
in line with this, the activation threshold is shifted towards higher input concentrations (i.e. smaller~$\hat{x}$) 
at the expense of reducing the dynamic range.
In the coupled system the optimal diffusion constant is markedly higher  ($\Delta^*=25$) than at $C=100$, 
and the resulting spatial averaging can almost completely remove the additional non-Poissonian noise contributions  ($\Phi(\hat x)\simeq 1$).
This allows the optimal activation threshold $K$ to shift towards  lower input concentrations in order to attain a similarly large dynamic range as for $C=100$.

At $C=0.01$, when input noise is exceedingly dominant, we observe the emergence of a qualitatively new regulatory strategy in the spatially coupled system.
For both spatially coupled and uncoupled systems, low $C$ favors sharp activation profiles with high Hill coefficients $H$. 
For the spatially uncoupled system, the optimal activation threshold $K^*$ moves towards central positions. 
In contrast, in the optimal spatially coupled system, the optimal activation threshold $K^*$ moves closer towards $\hat x=0$ where input noise is smaller, 
and the optimal Hill coefficient makes the activation nearly step-like. 
Without spatial coupling, this strategy could not be optimal, because it would be limited to at most one bit---in the region where the activation profile is flat, 
positions could not be discriminated based on the gene expression level $g$. 
Optimal diffusion can, however, smooth out this step-like activation  to generate a graded gene expression profile and restore position discriminability at essentially all positions. 
Simultaneously, optimal diffusion transforms the noise in an interesting way, as can be seen in Fig.~\ref{figOptimalProfiles}H and I. 
The step-like activation curve suppresses the super-Poissonian input noise everywhere except for the narrow interval of $x$ around the transition point 
where the super-Poisson component has a very sharp and tall peak; 
but this can very effectively be removed by the strong diffusive spatial averaging.
In sum, at low $C$, the optimal strategy is to use very sharp activation profiles, since they act as a ``filter'' for input noise (which must go to zero at zero or saturated gene expression); 
the strong optimal diffusion then smoothes out the step-like activation into a graded response and removes much of the remaining input noise around the very localized transition region at $c\approx K^*$. 

Usually, the effect of diffusion on information transmission is understood to be a tradeoff between the beneficial effect of noise averaging and detrimental effect of smoothing the profile, but here it seems that smoothing the step-like activation profile is also beneficial for the information flow. To check that this is indeed a qualitatively different optimal regulatory strategy, we performed a systematic study of the optimal regulatory parameters between $C=0.1$ and $C=0.01$. We found that here two regulatory strategies compete, as evidenced by two local maxima in the $(K,H)$ information planes for this range of $C$ values:
one favors relatively low Hill coefficients, $H^*\simeq 1$, and high activation thresholds $K^*$;
the other favors high $H^*$ and lower $K^*$.
For $C=0.01$ and $10 \lesssim \Delta \lesssim 30$ both regulatory strategies lead to almost equal information capacities,
but for $\Delta\gtrsim 30$ steep activation combined with fast diffusion is preferable.
Figure~\ref{figDoubleMax} in Appendix~\ref{secAppendixDoubleMax} illustrates this phenomenon.

\subsection{Full solution in 2D geometry differs only marginally from an approximate solution in 1D}
\label{secSpatialCouplingModelComparison}
All results presented in the previous sections were obtained using the 1D model with short-correlations assumption (SCA).
To assess how representative they are of the fully detailed model, we compared the 1D model with SCA
to the 2D model with SCA, and finally to the 2D model that retains more than only nearest-neighbor correlations
(see Appendix~\ref{secAppendixVarianceSolutions} for the formulas in the first two cases
and Appendix~\ref{secAppendixFullSolution} for the latter case).

Figure~\ref{figModelComparison} in Appendix~\ref{secAppendixModelComparison} compares the information planes $I(\Delta,\lambda)$ for the three cases at $C=1$. 
The results suggest that the influence of lateral spatial coupling present in the 2D model is minor.
Particularly in the range of spatial couplings $\Delta$ around the information maximum, where the system is operating close to the hard bound set by irreducible Poissonian noise,
the difference between information capacities in 1D and 2D is almost negligible.
Inclusion of 2D couplings mainly increases $I(x;g)$ values away from the optimal range of parameters,
thus effectively enlarging the region of parameter space in which the system can attain close-to-optimal information transmission.
This again is most prominent for the low $C$ regime, as expected, but still limited to a small fraction ($\lesssim 10\%$) 
of the total capacity (data not shown).
The effect of incorporating longer-ranged correlations is minor as well:
the information planes for the 2D model with and without SCA are almost indistinguishible.

Overall, this demonstrates that the 1D model with short-correlations assumption is a good approximation of the fully detailed model.
In the given context, the strength of spatial coupling appears more relevant than its topology.


\section{Discussion}
\label{secDiscussion}
We developed a generic framework to compute information flow through spatially coupled gene regulatory networks at steady state. 
As the simplest example, we considered how a spatially varying concentration field (``input profile'') is read out collectively by a regular 1D or 2D lattice of sampling units 
that are interacting by local diffusion of the gene product. 
A directly applicable example could be cells (or nuclei) in a developing multicellular organism that respond to the morphogen field by expressing developmental genes 
whose products can be exchanged between neighboring cells. 
We emphasize that our framework is generic in that  effects due to spatial coupling in any chosen geometry can be worked out before assuming any particular gene regulatory function 
and writing down the relevant noise power spectra, as is evident from Eqs.~(\ref{eqMeanSS}), (\ref{eqVarSS}) and (\ref{eqCorrSS}). 
This makes the framework widely applicable to a broad range of biological problems in which spatially distributed information is collectively 
sensed by a set of discrete sampling units and encoded into a spatial output.

The theory presented here is a direct continuation of our previous work 
on quantifying information flow in gene regulatory networks at increasing levels 
of realism \cite{Tkacik2008_PRE,Tkacik2009_Info,Walczak2010,Tkacik2012,Rieckh2014}. 
In our previous approaches, however, we were striving for analytical approximations that would permit computing the channel capacity, 
that is, performing analytical optimization over all possible distributions of input signals, $P_c(c)$; this led us to adopt the ``small noise approximation.'' 
Here, in contrast, we assumed a particular geometry (with a uniform distribution over sampling units, $P_x(x)$) and a particular functional form of the input concentration field, $c(x)$; 
the latter can depend on parameters which one can optimize, but we do not perform functional optimization over all possible $c(x)$. 
While these restrictions may appear stronger than in our previous work, they also allow us to move to a truly spatially discrete setup 
(that has a natural minimal spatial scale of a cell or nucleus, as is the case in reality), and permit to relax the small noise approximation, 
which is no longer assumed in this work. We also note that a parametric choice of $c(x)$ is not as restrictive as it might appear initially. 
First, one could choose a rich enough parametric form (basis set) for $c(x)$ that ensures spatial smoothness but otherwise allows optimization in a full space of plausible profiles. 
Second,  it is intuitively clear that monotonic functions encode positional information better than non-monotonic, or even constant functions, 
strongly narrowing the range of functional forms over which optimization should take place. 
Third, while we focused on exponential gradients which can be easily parametrized by only two parameters, exponential profiles 
actually are widespread throughout biology \cite{Kicheva2007,Wartlick2009,Bollenbach2008,DeLachapelle2010,Restrepo2014}.

Several previous approaches assessed how diffusive coupling alters the precision of spatial protein patterns 
driven by spatially distributed inputs \cite{Erdmann2009,Sokolowski2012,Holloway2011}.
While similar in spirit to ours, these works employed measures of ``regulatory precision'' such as the output pattern steepness and sharpness,
which---unlike mutual information---do not capture the problem in its full richness:
these measures quantify precision only locally, on parts of the output signal, and moreover, make implicit assumptions
about which feature of the output (boundary steepness, boundary position, etc.) is informative.
Information theory removes this arbitrariness \cite{Tkacik+Dubuis2015}, and permits extensions beyond the simple cases studied here. 
Similar approaches have recently been suggested for specific biological systems \cite{Mugler2014_qbio,Tikhonov2014_qbio,Taillefumier2014}.

In our example application we studied how spatial coupling alters optimal information flow
in a single-input/single-output gene regulatory network motivated by the \textit{bicoid-hunchback} system,
which is a part of the gap gene network in early fruit fly development \cite{Jaeger2011,Jaeger2012}.
The main outcome of this analysis is that diffusive coupling enhances information capacity
by removing super-Poissonian components of the noise, in line with previous work \cite{Erdmann2009,Sokolowski2012}; 
this effect is large when input noise is non-negligible ($C\leq 1$). 
When diffusion plays a large role in an optimal system, the activation functions can deviate substantially from the resulting output gene expression profiles. 
At the extreme, the activation functions can become step-like with very high Hill coefficients, while the gene expression profile still smoothly changes over its full dynamic range; 
this strategy, optimal at low $C$, where both spatial noise averaging and profile smoothing due to diffusion act jointly to increase information, 
is qualitatively different from the $C$ regime where profile smoothing is detrimental and trades off against noise averaging. 

In our setup, super-Poissoninan contributions to noise are fully accounted for by the input noise, but in reality this need not be the case. 
There are super-Poissonian contributions to noise that arise at the output, for instance, due to bursty production of proteins when the gene is activated. 
The simplest case is when mRNA expression is rate limiting, but then each mRNA can lead to a burst in the number of translated proteins. 
The theory can be simply extended to these cases by introducing a burst size into the relevant noise power spectra. 
Importantly, this would imply that diffusion can be effective at noise reduction (and thus beneficial for information transmission) also in the regime where $C\geq 1$. 
In support of this, recently another model linking mRNA expression and protein production with diffusion based on stochastic branching theory
has shown that deviations from Poissonian statistics in typical biological conditions 
are expected only for rather large burst sizes ($\gtrsim 50$) \cite{Cottrell2012}.

Recently, the field has made progress in detailed understanding of input-side noise in gene regulation due to stochastic diffusive arrivals 
of regulatory molecules to their binding sites \cite{Kaizu2014,Paijmans2014}. 
This has led to a revision of the previously suggested functional form for the Berg-Purcell limit \cite{Berg1977,Bialek2005} that we use in Eq.~(\ref{inoise}). 
Furthermore, we have also taken the simplest (Hill-type) regulatory function as our model for gene regulation, rather than picking a richer and potentially more realistic choice, 
such as the Monod-Wyman-Changeux form. 
These refinements will be included into our model in the future, but there is no reason to expect that they could change any qualitative outcome of our analyses.

It is interesting to speculate about the predictive power of optimization-based approaches applied to biological systems. 
In neuroscience, the principle of ``efficient coding'' which states that neural sensory systems devote their limited resources to maximize the information flow, in bits, 
from the naturalistic stimuli into the spiking neural representation \cite{Tkacik+Bialek2014_Review}, 
has proven enormously successful since the end of 1950s when it was first suggested by Barlow \cite{Barlow1961}. 
In signaling and gene regulation, similar ideas are much younger. On the other hand, the number of distinct genes or signaling proteins involved in the networks of interest 
is much smaller than the number of neurons; 
furthermore, molecular processes and the physical limits to sources of noise in gene regulation might be more easily understood than in neuroscience, 
where even a single neuron is a very complex object. 
It appears feasible that for small genetic networks the optimization problem 
would be tractable upon combining all phenomenology that until now we have analyzed separately: 
multiple, interacting genes, driven by one or more inputs, potentially with arbitrary feedback, all major relevant sources of noise, spatial coupling, 
and readout constraints \cite{Tkacik2008_PRE,Tkacik2009_Info,Walczak2010,Tkacik2012,Rieckh2014}. 
A major goal would also be the ability to relax the steady-state assumptions,
by considering either readout at particular time points (out of steady state), 
or the information between full state trajectories \cite{Tostevin2009,Tostevin2010,deRonde2010,Mugler2010,Selimkhanov2014}. 
Such a predictive theory, from which optimal networks could be derived mathematically, 
could then be realistically confronted with well-studied networks that can be quantitatively measured,
e.g., the gap gene network active during \Drosophila development \cite{Jaeger2011,Jaeger2012,Dubuis2013_MSB}.
It is intriguing to think that even in the molecular world the need to pay for abstract bits of information---in the currency of time or energy---led
nature to choose particular regulatory networks, and perhaps poised them at special operating points \cite{Krotov2014}.

\section{Acknowledgements}
The authors thank A. Mugler, I. Nemenman, T. Gregor, M. Tikhonov, P.R. ten Wolde and K. Kaizu for fruitful discussions.

\bibliography{Bib/GapGenes,Bib/InfoTransmission,Bib/LangevinEqn,Bib/General,Bib/SocialCells}

\clearpage
\newpage
\appendix

\setcounter{figure}{0}
\renewcommand{\thefigure}{S\arabic{figure}}

\section{Calculation of stationary means and (co)variances}
\label{secAppendixMeansVariances}
Considering Eq.~(\ref{eqODEmean}) in steady state immediately yields:
\begin{align}
 \lefteqn{ \la F(g_i,c_i) \ra = f(c_i) - \frac{1}{\tau} \la g_i \ra} 								\nn\\
 &= \hoprate \sum_{n_i\fromNNof{i}} \la g_i - g_{n_i} \ra = 2d\hoprate \la g_i \ra - h\sum_{n_i\fromNNof{i}} \la g_{n_i} \ra	\nn\\
\end{align}
This simply reflects the steady-state balance between production (and degradation) 
and diffusive in-/outflux in volume $i$.
Grouping $\la g_i \ra$ terms on one side gives
\begin{align}
 \la g_i \ra 			&= \frac{\tau}{1+2dh\tau} \left( f(c_i) + h\sum_{n_i\fromNNof{i}} \la g_{n_i} \ra \right)	\nn\\
 \Leftrightarrow\quad \mean_i 	&= T f(c_i) + \Lambda^2 \sum_{n_i\fromNNof{i}} \mean_{n_i}
\end{align}
where in the last step we abbreviate $\mean_i\equiv\la g_i \ra$, $T \equiv \tau/\left(1+2dh\tau\right)$, $\Lambda^2 \equiv hT$,
as in the main text.

To simplify the steady-state expression for the covariances resulting from Eq.~(\ref{eqODEvar})
it is instructive to treat its different parts separately.
For the terms correlating the fluctuations in volume $i$ with the production/degradation
process in volume $j$ we have (note that $\la \dg_i \ra = 0$ by construction):
\begin{align}
 \la \dg_i F(g_j,c_j) \ra &= \la \dg_i \ra f(c_j) - \frac{1}{\tau} \la \dg_i g_j \ra	\nn\\
			  &= - \frac{1}{\tau} \la \dg_i \left(\la g_j \ra + \dg_j \right) \ra = - \frac{1}{\tau} \la \dg_i \dg_j \ra
\end{align}
Hence,
\begin{align}
 \la \dg_i F(g_j,c_j) \ra + \la \dg_j F(g_i,c_i) \ra = - \frac{2}{\tau} \la \dg_i \dg_j \ra
\label{eqCorrProdDegr}
\end{align}
because $\la \dg_i \dg_j \ra \equiv \la \dg_j \dg_i \ra$.

Similarly, we can rewrite the term that correlates $\dg_i$ with the diffusive ``neighbor fluxes'' of $g_j$
\begin{align}
 \lefteqn{\hoprate \left\la \dg_i \sum_{n_j\fromNNof{j}} (g_{n_j} - g_j) \right\ra}							\nn\\
 &=\hoprate \left\la \dg_i \sum_{n_j\fromNNof{j}} \left( \la g_{n_j} \ra + \dg_{n_j} - \la g_j \ra - \dg_j \right) \right\ra		\nn\\
 &=\hoprate \left( \sum_{n_j\fromNNof{j}} \la \dg_i \ra \la g_{n_j} \ra + \la \dg_i \dg_{n_j} \ra \right) 				\nn\\
 &\qquad\qquad\qquad\qquad\qquad - 2d \hoprate \left( \la \dg_i \ra \la g_j \ra - \la \dg_i\dg_j \ra \right)				\nn\\
 &=\hoprate \left( \sum_{n_j\fromNNof{j}} \la \dg_i \dg_{n_j} \ra \right) - 2d\hoprate \la \dg_i \dg_j \ra
\label{eqCorrFluxes}
\end{align}
and analogously for the term in which $i$ and $j$ are exchanged.

Reinserting (\ref{eqCorrProdDegr}) and (\ref{eqCorrFluxes}) into the right side of Eq.~(\ref{eqODEvar}),
collecting covariance terms $\la \dg_i \dg_j \ra$ on the left side,
and multiplying by $\tau/2$
we obtain:
\begin{align}
 \lefteqn{\left( 1 + 2dh\tau \right) \la \dg_i \dg_j \ra}										\nn\\
 &= \frac{h\tau}{2} \left( \sum_{n_j\fromNNof{j}} \la \dg_i \dg_{n_j} \ra + \sum_{n_i\fromNNof{i}} \la \dg_j \dg_{n_i} \ra \right)	\nn\\
 &\qquad\qquad\qquad\qquad\qquad\qquad\quad + \frac{\tau}{2} \sum_k \left\la \Gamma_{ik}\Gamma_{jk} \right\ra				\nn\\
 \lefteqn{\Leftrightarrow \quad C_{ij} = \frac{\Lambda^2}{2} \left( \sum_{n_j\fromNNof{j}} C_{in_j} + \sum_{n_i\fromNNof{i}} C_{jn_i} \right)}	\nn\\
 &\qquad\qquad\qquad\qquad\qquad\qquad\quad + \frac{T}{2} \sum_k \left\la \Gamma_{ik}\Gamma_{jk} \right\ra
 \label{eqCovariancesGeneral}
\end{align}
where $C_{ij}\equiv\la\dg_i\dg_j\ra$.
The above equation couples the covariance $C_{ij}$ to the covariances $C_{in_j}$ and $C_{jn_i}$;
these represent the correlations between the protein number in volume $i$ and the neighbor volumes $n_j$ of $j$,
and the analogous quantity with $i$ and $j$ exchanged, respectively.
Hence, even if $i$ and $j$ are nearest neighbors on the lattice,
the expressions summing $C_{in_j}$ and $C_{jn_i}$ over $n_j$ and $n_i$, respectively,
will contain correlations between next-nearest neighbor volumes.
While the equation system defined by Eq. (\ref{eqCovariancesGeneral}) can be solved for the whole set
of covariances $C_{ij}$ upon imposing suitable boundary conditions,
this can be numerically expensive for larger spatial lattices and large parameter sweeps as part of optimization runs.
In this work, the quantity of interest is the variance $C_{ii}$ in volume $i$,
such that the calculation of longer-ranged correlations is not strictly required.

Considering the cases $i=j$ and $\nu\fromNNof{i}$ ($\nu$ being one of the nearest neighbors of $i$) 
in Eq. (\ref{eqCovariancesGeneral}) separately reveals the following interdependence between
the variance $\sigma^2_i\equiv C_{ii}$ and the nearest-neighbor covariance $C_{i\nu}$:
\begin{align}
 \sigma^2_i	&= \Lambda^2 \sum_{\nu\fromNNof{i}} C_{i\nu} + \frac{T}{2} \sum_k \left\la \Gamma_{ik}^2 \right\ra	\label{eqVarInterm}		\\
 C_{i\nu}	&= \frac{\Lambda^2}{2} \left( \sum_{n_\nu\fromNNof{\nu}} C_{in_\nu} + \sum_{n_i\fromNNof{i}} C_{\nu n_i} \right)			\nn\\
		&\qquad\qquad\qquad\qquad\qquad + \frac{T}{2} \sum_k \left\la \Gamma_{ik}\Gamma_{\nu k} \right\ra	\label{eqCovariancesNN}
\end{align}
Next-nearest correlations now only appear in Eq.~(\ref{eqCovariancesNN}),
which can be significantly simplified by an approximation
that we call the ``short-correlations assumption'' (SCA):
If the next-nearest-neighbor covariances are assumed to be small compared to the nearest-neighbor covariances 
and single-point variances, we can ignore them and set
\begin{align}
 C_{in_\nu}	= 0	\qquad \forall n_\nu \neq i	\nn\\
 C_{\nu n_i}	= 0	\qquad \forall n_i   \neq \nu
\end{align}
because then the only neighbor of $\nu$ that has some significant correlation with $i$ is $i$ itself, and vice versa.
In that case, the only remaining terms of the sums in Eq.~(\ref{eqCovariancesNN}) are $C_{ii}=\sigma^2_i$ and $C_{\nu\nu}=\sigma^2_\nu$, respectively:
\begin{align}
 C_{i\nu}	&= \frac{\Lambda^2}{2} \left( \sigma^2_{ii} + \sigma^2_{\nu\nu} \right)	 + \frac{T}{2} \sum_k \left\la \Gamma_{ik}\Gamma_{\nu k} \right\ra
 \label{eqCovariancesNN_SCA}
\end{align}
Specifying further the noise powers $\sum_k \left\la \Gamma_{ik}^2 \right\ra$ in (\ref{eqVarInterm})
and $\sum_k \left\la \Gamma_{ik}\Gamma_{\nu k} \right\ra$ in (\ref{eqCovariancesNN_SCA})
as described in Section~\ref{secMethodsMeansVariances} yields formulas (\ref{eqVarSS}) and (\ref{eqCorrSS}).

\section{Simplified variance formulae}
\label{secAppendixVarianceSolutions}
In our framework, computation of the mutual information $I(x;g)$ only requires the means $\mean_i$ and variances $\sigma^2_i$.
This enables us to reduce the number of coupled equations to be solved by direct insertion of (\ref{eqCorrSS}) into (\ref{eqVarSS}), 
thus eliminating $\cc_{ij}$.
For the 1D model with short-correlations assumption (SCA), after some algebraic steps this yields
\begin{align}
\sigma^2_i &= \frac{\Delta^2}{\left( 1+2\Delta \right)^2 - \Delta^2} \frac{1}{2} \left( \sigma^2_{i-1} + \sigma^2_{i-1} \right)	\nn\\
		&\quad + \frac{1}{\Nmax} \vast\lbrace \frac{1+2\Delta}{\left( 1+2\Delta \right)^2 - \Delta^2} \times			\nn\\
		&\qquad\qquad\qquad\quad \left( \frac{f(c_i) + \mean_i}{2} + \left[\left( \frac{\partial f}{\partial c} \right)^2 cc_0 \right]_{c_i} \right)	\nn\\
		&\qquad + \frac{\Delta}{\left( 1+2\Delta \right) + \Delta} \frac{1}{2} \left( \mean_{i-1} + 2\mean_i + \mean_{i+1} \right) \vast\rbrace
\label{eqVarianceSimplified1D}
\end{align}
where we have rewritten prefactor combinations containing $T$ and $\Lambda^2$
in terms of the spatial coupling $\Delta$.

The analogous formula for the 2D model with SCA reads
\begin{align}
 \sigma^2_{(ij)} &= \frac{\Delta^2}{\left( 1+4\Delta \right)^2 - 2\Delta^2} \frac{1}{2} \sum_{n_{(ij)}} \sigma^2_{n_{(ij)}}	\nn\\
		&\quad + \frac{1}{\Nmax} \vast\lbrace \frac{1+4\Delta}{\left( 1+4\Delta \right)^2 - 2 \Delta^2} 		\nn\\
		    &\qquad\qquad\quad \times \left( \frac{f(c_{(ij)}) + \mean_{(ij)}}{2} + \left[\left( \frac{\partial f}{\partial c} \right)^2 cc_0 \right]_{c_{(ij)}} \right)					\nn\\
		&\qquad + \frac{\Delta \left( 1 + 3\Delta \right)}{\left( 1+4\Delta \right)^2 - 2 \Delta^2} \frac{1}{2} \left[ 4\mean_{(ij)} + \sum_{n_{(ij)}} \mean_{n_{(ij)}} \right] \vast\rbrace	\nn\\
\label{eqVarianceSimplified2D}
\end{align}
where we indicate volumes of the two-dimensional lattice with a 2D index $(ij)$,
and the $n_{(ij)}$-sums run over the four nearest neighbor volumes of $(ij)$.

\section{Full solution without short-correlations assumption}
\label{secAppendixFullSolution}
For completeness,
here we also state the formula defining the linear system for the coupled covariances
in the full two-dimensional model without short-correlations assumption,
again indicating volumes by the two-dimensional index $(ij)$:
\begin{align}
 \cc_{(ij)(kl)} &= \frac{1}{2}\frac{1}{1+4\Delta} \Bigg\lbrace \hat\Noise^2_{(ij)(kl)}			\nn\\ 
		&\qquad\qquad + \Delta\Big[\quad 	 \cc_{(ij)(k-1,l)} + \cc_{(ij)(k+1,l)}		\nn\\
		&\phantom{\qquad\qquad\Delta\Big[\quad} + \cc_{(ij)(k,l-1)} + \cc_{(ij)(k,l+1)}		\nn\\
		&\phantom{\qquad\qquad\Delta\Big[\quad} + \cc_{(i-1,j)(kl)} + \cc_{(i+1,j)(kl)}		\nn\\
		&\phantom{\qquad\qquad\Delta\Big[\quad} + \cc_{(i,j-1)(kl)} + \cc_{(i,j+1)(kl)} 	\Big]\Bigg\rbrace	\nn\\
\label{eqCovariancesNoSCA}
\end{align}
where the normalized noise term $\hat\Noise^2_{(ij)(kl)}$ is only non-zero 
if the volumes indicated by $(ij)$ and $(kl)$ are identical, i.e. $(ij)\equiv(kl)$, 
or nearest neighbors, i.e. $(kl) \in N((ij))$,
and takes one of the following forms, respectively:
\begin{align}
 \lefteqn{ \hat\Noise^2_{(ij)(kl)} = \hat\Noise^2_{(ij)(ij)} }	\nn\\
 && = \hat\GNoise^2_{(ij)} + \sum_{n_{(ij)}} \left( \hat\DNoise^2_{[(ij)\rightarrow n_{(ij)}]} + \hat\DNoise^2_{[n_{(ij)}\rightarrow(ij)]} \right)	\nn\\
 &&\text{if } (ij) \equiv (kl)				\nn\\
 && 							\nn\\
 \lefteqn{ \hat\Noise^2_{(ij)(kl)} = - \left( \hat\DNoise^2_{[(ij)\rightarrow(kl)]} + \hat\DNoise^2_{[(kl)\rightarrow(ij)]} \right) }			\nn\\
 && 							\nn\\
 &&\text{if } (kl) \in N((ij))				\nn\\
 && 							\nn\\
 \lefteqn{ \hat\Noise^2_{(ij)(kl)} \equiv 0 \qquad\qquad\qquad\qquad\, \text{else} }
\end{align}
Here $n_{(ij)}$ runs over the four nearest-neighbors of volume $(ij)$.
The normalized noise powers $\hat\GNoise^2_{(ij)}$ and $\hat\DNoise^2_{[(ij)\rightarrow(kl)]}$
derive from the expressions presented in Section~\ref{secMethodsNoiseAndRegF}:
\begin{align}
 &\hat\GNoise^2_{(ij)} = \frac{1}{\Nmax} \left( f(c_{(ij)}) + \mean_{(ij)} + \left[\left( \frac{\partial f}{\partial c} \right)^2 2cc_0 \right]_{c_{(ij)}} \right)	\nn\\
 &\hat\DNoise^2_{[(ij)\rightarrow(kl)]} = \frac{\Delta}{\Nmax} \mean_{(ij)}
\label{eqNoisePowersFullModel}
\end{align}
Eq. (\ref{eqCovariancesNoSCA}) in principle allows for calculation
of covariances between volumes that are arbitrarily far apart.
In practice, because of the finiteness of the lattice, the spatial correlations 
still have to be truncated at a certain distance and boundary conditions applied 
in order to obtain a closed system that consists of as many equations as variables.

\section{Standard parameters}
\label{secAppendixParameters}
\begin{table}
\begin{tabular}{|l|c|l|}
 \hline
 {\bf Quantity}				& {\bf Symbol}	& {\bf Value}		\\
 \hline
 Lattice spacing 			& $\delta$	& 8.33~$\mu m$			\\
 No. of nuclei in $x$-direction		& $N_x$		& 60				\\
 -- resulting system length		& $L$		& 500~$\mu m$			\\
 \hline
 Internal activator diffusion constant	& $D_c$		& 3.16~$\mu m^2 / s$		\\
 Activator binding site length		& $l_c$		& 0.01~$\mu m$			\\
 Product protein lifetime		& $\tau$	& 240~$s$			\\
 Std. max. mean product copy no.	& $\Nmax^0$	& 444				\\
 -- resulting typ. concentration	& $c_0$		& 58.5~$\mu m^{-3}$		\\
					&		& ($\simeq35~nM$)		\\
 -- resulting typ. diffusion constant	& $D_0$		& 0.29~$\mu m^2 / s$		\\
 \hline
 Std. input gradient length		& $\gamma_0$	& 100~$\mu m$			\\
 Std. input gradient amplitude		& $\cmax$	& $=c_0$			\\
 \hline
\end{tabular}
\caption{The standard parameter values of our model.}
\label{TabParameters}
\end{table}

In our derivations most system parameters can be combined into two natural scales,
a typical concentration $c_0$ (cf. Section~\ref{secMethodsNoiseAndRegF}) and a typical diffusion constant $D_0$ (cf. Section~\ref{secMethodsMeansVariances}),
and we measure concentrations and diffusion in these scales throughout our theory.
To obtain numerical solutions, however, concrete numbers have to be assigned to these quantities.

Since our example application represents the \textit{bicoid-hunchback} system in early \Drosophila development,
we opted for the following choice for the baseline values of the parameters in Table~\ref{TabParameters}:
First we chose lattice parameters that roughly correspond to the geometry of the \Drosophila syncytium \cite{Gregor2007a,Gregor2007b};
in particular, this defines the lattice spacing $\delta$.
We then chose a typical value for the product protein lifetime $\tau$, which defines the typical diffusion scale $D_0=\delta^2/\tau$.

To set the value for $c_0$, we took advantage of the fact that in the \textit{bicoid-hunchback} system the activator (i.e. Bicoid) concentration at midembryo
has been measured experimentally \cite{Houchmandzadeh2002,Gregor2007a}.
This lead us to set $c_0$ equal to the extrapolated maximal value of the measured in-vivo gradient at its source ($x=0$).
In other words, at $C=\cmax/c_0=1$ the input function $c(x)$ in our model corresponds to the experimentally measured Bicoid gradient.
Finally, to define a standard value for the maximal mean output $\Nmax$, which is a key determinant of the noise powers and 
variances (cf. Eqs.~(\ref{eqNormNoisePowerCov}), (\ref{eqNormNoisePowerVar}), (\ref{eqVarianceSimplified1D}), (\ref{eqVarianceSimplified2D}) and (\ref{eqNoisePowersFullModel}))
but to-date unknown,
we chose typical values for the internal activator diffusion constant $D_c$ and the binding site length $l_c$;
from this, we calculated the standard value $\Nmax^0=c_0 D_c l_c \tau$.
When varying $C$ away from the baseline setting, we changed either $\cmax$ while holding $\Nmax$ (and thus $c_0$) constant,
or by varying $\Nmax$ and keeping $\cmax$ unchanged, as described in Appendix~\ref{secAppendixVaryNmax}.

Table~\ref{TabParameters} demonstrates that all baseline parameter values are well within a biologically realistic regime.

\pagebreak

\section{Varying $C$ via $\Nmax$}
\label{secAppendixVaryNmax}
\begin{figure}[t]
\begin{center}
  \includegraphics[width=\linewidth]{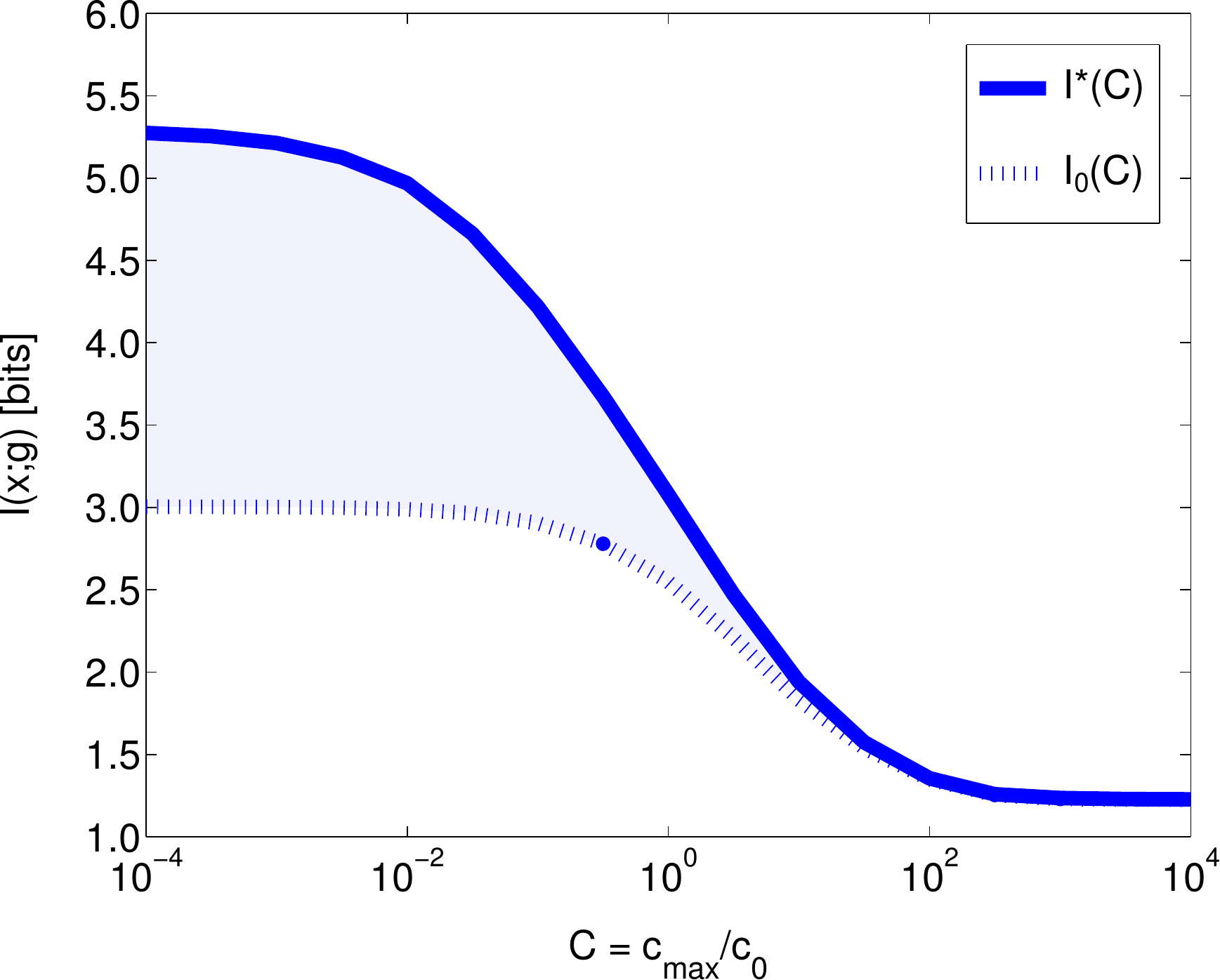}
\end{center}
\caption{{\bf Information capacity for varying noise type ratios, varied via $\Nmax$.}
	 Shown is the optimized information capacity as a function of the noise type ratio $C=\cmax/c_0$
	 with optimal diffusive coupling ($I^*(C)$, solid blue line)
	 and without diffusive coupling ($I_{0}(C)$, dashed blue line).
	 $C$ was varied via $\Nmax$.
	 At small $C$ values, non-Poissonian input noise is dominant, while Poissonian output noise dominates for large $C$.
	 The blue-shaded area depicts the maximal gain in information capacity from diffusive coupling.
}
\label{figImaxVsC_Nmax}
\end{figure}
In our model, the noise type ratio $C = \cmax/c_0 = \cmax / \Nmax \cdot (D_c l_c \tau)$ can be varied in multiple ways.
In addition to altering the importance of non-Poissonian input noise via $\cmax$, 
we also studied the case in which the contribution of Poissonian output noise is varied via $\Nmax$ while $\cmax$ is held constant.

The main difference to the case in which $\cmax$ is scaled is that now information capacities decrease with increasing $C$, 
as this means decreasing $\Nmax$ and thus enhancing output noise.
The highest information capacity then is attained in the limit $C\rightarrow 0$, i.e. $\Nmax \rightarrow \infty$.
In the uncoupled system, $I(x;g)$ saturates for $C\rightarrow 0$ towards a value 
dictated by the amount of (in this case) irreducible input noise, set by $\cmax$.
With spatial coupling, optimal $I(x;g)$ values in the low-$C$ regime are markedly higher,    
in accordance with the finding that spatial averaging can only enhance information capacity when input noise is dominant.
As expected, for $C\rightarrow 0$, i.e. negligible output noise, and sufficiently strong spatial coupling, i.e. strongly attenuated input noise,
the capacity approaches the hard bound of the noise-free limit
given by the finite number of sampling points $N_x$ along the $x$-axis, $I_{N_x}^{\rm max}=\log_2(N_x)$.

\section{Double maximum}
\label{secAppendixDoubleMax}
\begin{figure*}
\begin{center}
    \includegraphics[width=\linewidth]{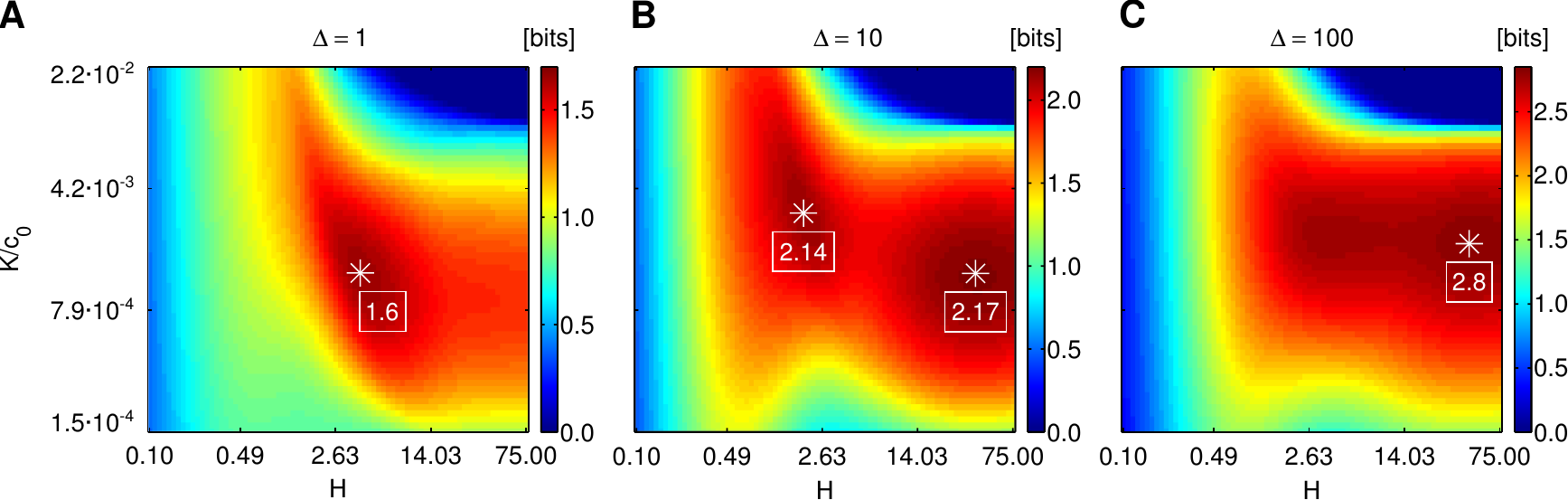}
\end{center}
\caption{{\bf Emergence of a second maximum in the information plane.}
	  Here we plot for $C=0.01$ and standard input gradient length, i.e. $\lambda=1$,
	  the the positional information $I(x;g)$ 
	  as a function of the regulatory parameters $H$ and $K$
	  for increasing strength of spatial coupling:
	  {\bf (A)} $\Delta=1$, {\bf (B)} $\Delta=10$, and {\bf (C)} $\Delta=100$.
	  White stars mark local maxima,
	  white boxes show the corresponding amount of information in bits.
	  The data shown is for the 2D model with SCA.
}
\label{figDoubleMax}
\end{figure*}

In Fig.~\ref{figDoubleMax} we show three different information planes
for noise-type ratio $C=0.01$ and increasing values of the spatial coupling $\Delta$ ($\Delta=1$, $\Delta=10$ and $\Delta=100$).
The heat maps demonstrate the emergence of distinct optimal regulatory strategies for $\Delta>1$:
for sufficiently strong diffusion ($\Delta=10$), very steep activation curves ($H\gg 1$) 
at lower activation thresholds $K$ result in similar information transmission
as less steep activation curves ($H\simeq 1$) at slightly higher $K$.
For $\Delta=100$, steep activation performs better than the other strategy.
This effect is seen most clearly in the 2D model (with SCA),
but also appears in the 1D model at slightly different values of $C$ and $\Delta$.

\section{Model comparison}
\label{secAppendixModelComparison}
\begin{figure*}
\begin{center}
   \includegraphics[width=\textwidth]{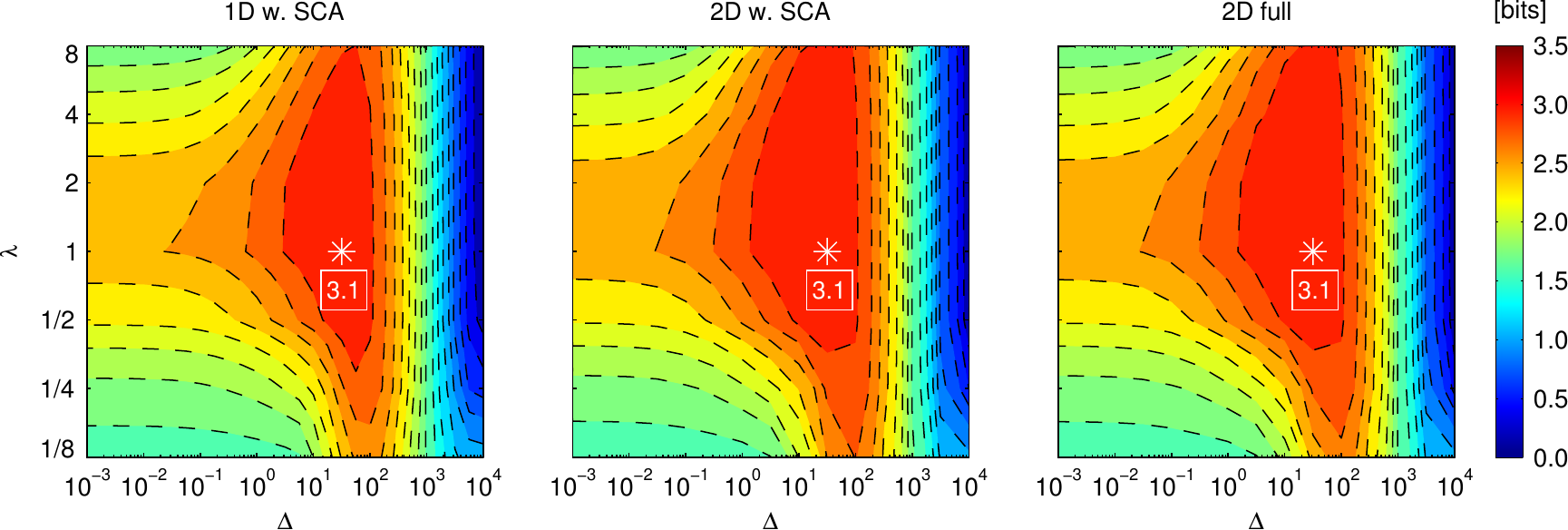}
\end{center}
\caption{{\bf Comparison of $I(x;g)$ for increasingly detailed versions of the spatial-stochastic model at $C=1$.}
	  The reference plot for the 1D model with SCA (left) is identical to the information plane for $C=1$ in Fig.~\ref{figOptimumVsDAndLF}.
	  The same information plane is shown for the 2D model with SCA (middle),
	  and for the full 2D model that retains next-nearest neighbor correlations (right).
	  }
\label{figModelComparison}
\end{figure*}
In Fig.~\ref{figModelComparison} we demonstrate how including increasing detail
affects our results for a paradigmatic case ($C=1$, $\lambda=1$).
The different panels show the same information plane for
the 1D model with SCA, the 2D model with SCA and
the 2D model without SCA, which retains longer-ranged correlations
with next-nearest neighbor volumes.

\end{document}